\documentclass[aps,preprint,prb,eqsecnum]{revtex4}
\usepackage{epsfig,graphics,latexsym}
\newcommand{\vn}{{\bf n}}
\newcommand{\vx}{{\bf x}}
\newcommand{\vy}{{\bf y}}
\newcommand{\vR}{{\bf{R}}}
\newcommand{\vq}{{\bf q}}
\newcommand{\vg}{{\bf g}}

\newcommand{\vG}{{\bf{G}}}
\newcommand{\bea}{\begin{eqnarray}}
\newcommand{\eea}{\end{eqnarray}}
\newcommand{\be}{\begin{equation}}
\newcommand{\ee}{\end{equation}}
\newcommand{\nn}{\nonumber}

\newcommand{\wbar}{\bar{w}}
\newcommand{\Fp}{F^{\prime}}

\begin{document}
\title{Quantum Hall Skyrmion Lattices at $g \approx 0$} 

\author{Sumithra Sankararaman\footnote{Email:sumithra@imsc.res.in}
and R. Shankar\footnote{Email:shankar@imsc.res.in}}
\affiliation{The Institute of Mathematical Sciences,\\
C.I.T. Campus, Taramani, Chennai 600 113,
India.}

\begin{abstract}
  We investigate the classical ground state configurations of a
  collection of skyrmions in the limit of vanishing L\'ande $g$
  factor. We show that in this limit, the skyrmions which are large
  and overlapping, behave qualitatively differently from small
  skyrmions. We investigate the system in various regimes of spin
  stiffness. We find that in all regimes, at $g=0$, the ground state
  configuration is a rectangular skyrmion lattice with N\'eel
  orientational ordering. The charge distribution has a higher
  symmetry of a face-centered rectangular lattice and the
  $z$-component of the spin polarization is zero. We also present
  general argument for the vanishing of spin polarization at $g=0$ at
  finite temperature and in the presence of disorder and discuss
  some experimental consequences of our results. 
\end{abstract}

\pacs {PACS : 73.43.-f,73.43.Cd,73.43.Lp} 

\date{\today}
\pagebreak
\maketitle

\section{Introduction}

The lowest energy charged excitations about the $\nu=1$ ferromagnetic
quantum Hall ground state are skyrmions. They are topological objects
in which the spin gradually twists over an extended region. Their spin
is greater than $1/2$ and they carry an electric charge of $\pm e$.
Skyrmions have a topological charge which equals their electric charge
at $\nu=1$ \cite{sondhi,lee}. Skyrmionic excitations are favoured over
single particle excitations when the Land\'e g-factor is small i.e $g
\rightarrow 0$ \cite{sondhi,fertig1,fertig2}. Skyrmions which are
produced around the $\nu=1$ ground state have been experimentally seen
by in OPNMR and optical magneto-absorption experiments on Ga nuclei in
an electron doped multiple quantum well structure
\cite{barrett,aifer}. A sharp fall in the spin polarisation of the 2D
electron gas is seen on either side of $\nu=1$, which indicates that a
number of spins are being flipped by the addition or removal of a
single electron, in contrast to single particle excitations where no
extra spin is flipped. Small skyrmions have also been observed as low
energy excitations about the $\nu=1$ state in tilted field
measurements reported in \cite{eisenstein}. The above experiments
observe skyrmions where the number of flipped spins varies between $3$
and $7$. Low $g$ experiments observe large skyrmions with a large
number of flipped spins ($50$ and $36$ respectively)
\cite{shukla,nick}. In these experiments a low value of $g$ is
obtained either by suitably altering the stoichiometry while growing
the sample or by application of hydrostatic pressure.

The low temperature ground state of a system of interacting skyrmions
is expected to be a crystalline lattice \cite {brey,green,madan,timm}.
Earlier calculations performed in the mean field limit by Brey {\it
  et. al.} \cite{brey} suggest that the ground state of a system of a
skyrmions is a square Skyrme crystal with an anti-ferromagnetic order
in the planar component of the spin of the skyrmion. Later studies by
Green {\it et. al.} , Rao {\it et. al.} and Timm {\it et. al.} analyze
the classical ground state of the skyrmion lattice as $\nu \rightarrow
1$ using the non-linear sigma model (NLSM) \cite{green,madan,timm}.
Green {\it et. al} assume that the ground state is a triangular
lattice with spiral ordering \cite{green}.  Rao {\it et. al.} report
$triangle \rightarrow square \rightarrow triangle$ transitions in the
skyrmion lattice as a function of the filling factor near $\nu = 1$
and at T=0 \cite{madan}. Timm {\it et. al.} study a system of well
separated skyrmions as described by an anti-ferromagnetic XY model and
propose a T=0 phase diagram in which the triangle and square phases
are separated by N\'eel ordered, face-centered rectangular phases
\cite{timm}. The classical and quantum phase transitions occurring in
the Skyrme crystal has been studied using Hartree-Fock calculations by
C\^ot\'e {\it et. al.} \cite{stoof}. Moon {\it et. al.} study the
renormalization effect of skyrmions and the thermodynamics of a
skyrmion system, without assuming apriori lattice structures
\cite{moon}.

At non-zero $g$, the charge and spin densities of skyrmions are peaked
about its north pole if the vacuum spin polarization direction is
taken to be the south pole. They fall off exponentially from their
central value.  The corresponding size is determined by the
competition between the Coulomb repulsive energy and the Zeeman
energy. The former favours an infinite size (power law fall-off) and
the latter a zero size. The $XY$ component of the skyrmion spin
density assumes a "hedgehog" configuration. These configurations can
be characterized by the orientation angle subtended between the
position vector (with respect to the center of the skyrmion) and the
$XY$ component of the spin density at that point. The finite sized
skyrmion therefore behaves like an oriented, charged disc. It is
energetically favourable for nearby skyrmions to have opposite
orientations. These considerations lead to small skyrmions being
modeled by discs carrying an $XY$ spin with coulomb repulsion and
short range anti-ferromagnetic interactions \cite{timm}. The coulomb
repulsion leads to a triangular lattice in the dilute limit and the
$XY$ antiferromagnetism in the orientational degree of freedom to the
square lattice as the inter-skyrmion spacing decreases.

In this paper we show that this picture breaks down when the
skyrmionic size is larger than the inter-skyrmion spacing. For
instance, when $g=0$ there is no exponential damping and we show that
the charge density is symmetric in the neighbourhood of the north and
south poles. The oriented disc picture is then clearly invalid. We
study the ground state configuration of a collection of skyrmions at
$g=0$ and $g=0.05g_0$, ($g_0$ is the physical value of the Land\'e $g$
factor). The energy of the system at $g \approx 0$ is described well
by the Non-linear Sigma Model energy functional
\cite{numericalsondhi}.  We minimize this energy functional in three
different regimes: (1) High stiffness regime in which the gradient
term dominates the energy functional, (2) Low stiffness regime in
which the Coulomb term dominates and (3) Intermediate stiffness in
which both terms in the energy functional compete to generate a
suitable ground state. Finally, we discuss the behaviour of the spin
polarisation and its experimental consequences.

The rest of the paper is organised as follows. Section II, describes
the NLSM energy functional. Section III discusses the ansatz used to
solve the energy functional and the symmetries of that ansatz. In
Section IV we discuss the minimum energy configuration in the high
stiffness ,low stiffness and intermediate stiffness regimes.  Section
V we describe a general argument for the vanishing of the spin
polarization at $g=0$ and discuss the experimental consequences.
Finally, the conclusions of this paper and a summary of the results is
presented in Section VI.

\section{Model and NLSM energy functional}
We consider skyrmionic excitations about the $\nu=1$ ground state. The
number density of skyrmions is given by $n_{sky} =
\frac{1-\nu}{\nu}n_c$, where $n_c$ is the carrier density. In our
calculations we use the carrier density, $n_c = 1.5 \times
10^{11}~cm^{-2}$ and change $\nu$ by tilting the magnetic field. We
localize the skyrmion centers on the lattice points of a Bravais
lattice shown in Fig.\ 1.  We choose a skyrmion density ($n_{sky}$) of
1 skyrmion per unit cell (i.e one skyrmion per lattice point).

We measure all lengths in units of the magnetic length $l_{c}=\sqrt
\frac {\hbar} {e B}$ and all energies in terms of the cyclotron energy
$\hbar \omega_c$, where $\omega_c = (eB/m^*c)$, $m^*$ is the effective
mass of the electron.  The area of the unit cell is fixed by the
filling factor and is given by

\be
A_{\Box} = \frac {1}{n_{sky}} = \frac{2\pi}{1-\nu}
\ee

The local spin polarisation which is represented by unit vector field
$\vn(\vx)$ is stereographically projected onto the complex plane by
the transformation $w=\cot(\theta/2) e^{i\phi}$, where $\theta$ and
$\phi$ are the polar angles of the spin vector $\vn (\vx)$. In the
rest of the paper we will work with the planar spin variable $w$.
 
The topological charge density is given by
\bea
\rho(\vx) &=& \frac{1}{4\pi}\epsilon_{ij}\vn.(\partial_i \vn \times 
\partial_j \vn) \nn \\
&=& \frac {\epsilon_{ij}
\partial_i w \partial_j \overline w} {2 \pi i (1+w \overline w)^{2}}
\eea
The topological charge, 
$Q(\vx)=\int_{\Box} d^{2}x~\rho(\vx) = 1$
($\int_{\Box}$ denotes integration over the unit cell).

The low energy, long wavelength excitations about the $\nu=1$ ground
state are described by the NLSM.  The NLSM energy functional has to be
minimized for different filling factors to get the minimum energy
configurations of this lattice.  The NLSM energy functional with
Zeeman and Coulomb interactions is \cite{sondhi,green,madan}:

\be
E = E_{grn}+  E_z + E_{coul}
\ee

The gradient or the spin exchange term proportional to $\int d^2x
{\mid \partial_i \vn(\vx) \mid }^2,(i = x,y)$, is calculated as

\be
\label{grn}
E_{grn} = \frac{\gamma}{2} \int_{\Box} d^{2}x \frac {(\partial_x w
  \partial_x \overline w +\partial_y w \partial_y \overline w )} {(1 +
  w \overline w)^2} 
\ee 

where $\gamma = \frac {e^*}{16 \sqrt{2 \pi}}$.  The gradient energy
density is {\it $ E_{grn} /A_{\Box}$} , where $A_{\Box}$ is the area
of the unit cell.  This term alone is the pure NLSM and it has exact
scale invariant solutions \cite{poly}.

The Zeeman term is proportional to the z-component of the
total spin i.e. to
$\frac{\nu}{2 \pi} \int_{\Box} d^2x
\frac {(1 + n^z)}{2}$ (where $n^z$ is the z-component of $\vn(\vx)$).
In our units the average number of electrons is $\frac{\nu}{2
\pi}$. 
The z-component of the total spin is 
\bea
(Total~spin)_z =  \frac{\nu}{2 \pi}\int_{\Box} d^{2}x \frac 
{\overline w w}{(1 + \overline w w)}
\eea
Therefore,
\bea
\label {ezee}
E_z = g^* \frac{\nu}{2 \pi}\int_{\Box} d^{2}x \frac {\overline w w}
{(1 + \overline w w)}
\eea
where $g^{*} =\frac {g \mu_B B}{\hbar \omega_c}$.
The Zeeman energy density is therefore {\it $E_z/A_{\Box}$},where $A_{\Box}$ is the 
area of the unit cell.

The Coulomb energy density term is a term of the form
\be
E_{coul} = \frac{e^*}{2}\frac {1}{A_{tot}} \int_{\vx,\vy} \rho(\vx)
\frac {1}{\mid \vx - \vy \mid} \rho(\vy)
\ee
where $e^{*} = (e^2/Kl_c)(1/\hbar \omega_c)$ and $A_{tot}$ is the 
total area of the lattice. 

The Coulomb term arises because the electric charge density is 
proportional to the topological charge density. Since the topological 
charge density explicitly appears in the above expression the 
spin orientation gets automatically tied to the Coulomb energy. 
The four dimensional integral in the Coulomb term can be converted 
to a sum over the reciprocal lattice.
\bea
E_{coul} &=& \frac{1}{A_{\Box}} \sum_{\{\vR\}} \int_{\vx,\vy \in \Box} \rho(\vx) \frac
{1}{\mid \vx-\vy+\vR \mid} \rho(\vy) \nn \\
&=& \frac{1}{A_{\Box}} \sum_{\vR} \int \frac{d^2 g}{(2 \pi)^2} e^{i \vg.\vR} 
{\mid \tilde{\rho} (\vg) \mid}^2 \frac{2 \pi} {\mid \vg \mid}
\eea    
After doing the summation over $\{\vR\}$ and converting the integral over $\vg$ to
a sum we get
\be
{E_{coul} =  \frac {e^{*} \pi}{A_{\Box}^2} \sum_{\{\vG_R\}} {\mid \tilde{\rho}
(\vG_{R})\mid}^{2}
\frac {1} {\mid \vG_R \mid}}
\ee
where $\tilde{\rho} (\vG_{R}) = \int_{\vx \in \Box} \rho(\vx) e^{i \vG_{R}.
\vx}$, 
$\vG_R$ lies in the reciprocal lattice and
$A_{\Box}$ is the area of the unit cell.

\section{Ansatz and Symmetries }

In this section we will describe and motivate the ansatz we will use
to minimise the energy functional discussed above. Let us first
consider the model without the Coulomb and Zeeman interactions. In
this case it is known \cite{poly} that any meromorphic function $w(z)$
is an exact solution of the equations of motion. Since the positions
of the poles and zeros define a meromorphic function upto a
multiplicative constant, the most general $N$ skyrmion solution can be
written as,

\begin{equation}
\label{nsky1}
w_0(z)=w_\infty\frac{\prod_{m=1}^N (z-{\bar \vR}_m)}{\prod_{n=1}^N(z-\vR_n)}
\end{equation}

$w_\infty$ specifies the direction of the spin polarization at $\vert
z\vert \rightarrow \infty$. Equation (\ref{nsky1}) can be rewritten
as,

\begin{equation}
\label{nsky2}
w_0(z) = w_\infty+\sum_{n=1}^N \frac{\lambda_n}{(z-\vR_n)}
\end{equation}
$\lambda_n$ are $N$ complex numbers which are given by the solution of the
$N$ linear equations,
\begin{equation}
\label{lambdas}
w_\infty+\sum_{n=1}^\infty \frac{\lambda_n}{{\bar \vR_m}-\vR_n}=0
\end{equation}

If we choose $w_\infty=0$, corresponding to the spin polarization at
infinity pointing in the $-\hat z$ direction, and take the $N
\rightarrow \infty$ limit, equation (\ref{nsky1}) is written as,

\begin{equation}
\label{w0}
w_0(z) = \sum_{\vR} \frac{\lambda (\vR)}{(z-\vR)}
\end{equation}

This motivates an ansatz of the form
\be
\label{latans}
w(z)  =  w_0(z) F(z)
\ee
for the fully interacting model. The skyrmion lattice is obtained by
letting $\{\vR\}$ go over the position vectors of a lattice. For a
Bravais lattice with basis vectors $e1$ and $e2$, $\vR = n e1+m e2$,
where $n$ and $m$ are integers. Further when $|e1|=|e2|=e$ these basis vectors
describe a face-centered rectangular lattice as shown in Fig.\ 1.
They subtend an angle $\theta$ at the origin.  When $\theta = 2 \pi/3$ the
lattice is a triangular lattice.
When $\theta = \pi/2$ the lattice is rectangular with $a$ and $b$
as the length of its sides. $a/b$ is the aspect ratio of the rectangle.
$a/b = 1$ corresponds to the special case of a square lattice.
$F(z)$ is a smooth, non-analytic function with the periodicity of the
lattice. For the rectangular lattice we take it to be of the form,
\bea
\label{fz}
F(z) &=& \frac{e^{-\kappa_{np} r_{np}}}{e^{-\kappa_{sp} r_{sp}}} \nn \\
r_{np} &=& \sqrt{\sin^2(\pi x/a)+\sin^2(\pi y/b)} \nn \\
r_{sp} &=& \sqrt{\cos^2(\pi x/a)+\cos^2(\pi y/b)} \nn
\eea

The spin configuration corresponding to this ansatz is quasi-periodic. 
We have,
\be
w(z+{\vR})=e^{i\vq.\vR}w(z)
\ee
i.e. the spin rotates about the $z-$axis by and angle $\vq.\vR$ when 
translated by $\vR$. However the energy has the periodicity
of the lattice (since all the energies depend on $\overline{w}w$.
As shown in earlier work \cite{madan}, $w(z)$ describes a configuration
with skyrmion number one per lattice site. The mapping $w = cot(\theta/2) 
e^{i \phi}$ implies that the anti-podal points $\theta = 0$ and $\theta = \pi$ 
correspond to the poles and zeroes of $w_0(z)$. In the rest of the paper, 
we will refer to the poles as the ``north poles'' and the zeroes as the 
``south poles'' respectively. 

If we take $\kappa_{sp}=0$ and $1/\kappa_{np} << a,b$, then the ansatz
in equation (\ref{latans}) is essentially the same as the one used in
\cite{madan}. The charge and spin densities are then concentrated
about the north poles. In this regime, each skyrmion can be visualised
as a charged, oriented disc. The orientation corresponding to the
angle the $XY$ component of the spin density at any point makes with
the vector joining that point with the center of the disc. The ansatz
in equation (\ref{latans}) then corresponds to a collection of well
separated skyrmions centered at the lattice sites (see Fig.\ 1). The
vector $\vq$ can be interpreted as specifying a spiral orientation
order for these skyrmions.  The orientation rotates by $\vq.\vR$ when
we move by $\vR$. This is the picture used in reference \cite{timm}
and the classical phase diagram can be qualitatively understood in
terms of the long range coulomb repulsion between the discs (which
favours a triangular lattice) and a short range
antiferro-orientational interactions (which prefer an unfrustrated
square lattice).  When $g=-0.44$, the value for $GaAs$ at low
pressures, previous theoretical and experimental work show that the
skyrmions have a spin $s\approx 5$.  The spin can be related to the
rms charge radius as $\langle r^2 \rangle = 2s$ (see Appendix 2).
Skyrmions can be called small when this radius is smaller than half
the lattice spacing $\sqrt{\frac {2 \pi}{(1-\nu)}}$. Skyrmions of spin
$\approx 5$ have a size of $\approx 3$.  At $\nu = 0.9$, the lattice
spacing is around $8$.  The skyrmions are thus well separated at $\nu
\ge 0.9$.  The system of skyrmions in this regime is well described by
the oriented, charged disc picture. As $g$ is decreased, the skyrmions
become larger and, as we will show below, this picture breaks down.

We show plots of the spin density and charge density in a lattice of
well separated skyrmions. Figures 2. and 3. show the XY spin
configuration and the charge density in a triangular lattice of well
separated skyrmions \cite{color}.  The spins are ordered into a triangular lattice
(with ABC sub-lattice structure) with $\vq = (2 \pi/3,2 \pi/3)$.
Figures 4. and 5. show the XY spin configuration and the charge
density distribution for a square lattice of well separated skyrmions.
The spins have N\'eel order (with AB sub-lattice structure) with $\vq
= (\pi,\pi)$.  We notice that the charge is peaked only around the
``north poles'' (the lattice points).

Let us now consider the extreme case i.e. $g=0$. We first look at the
high stiffness limit i.e. when the coefficient of the gradient term,
$\gamma$ is large. The gradient energy then dominates the NLSM energy
functional. This energy is exactly minimised by any analytic ansatz.
So in this regime, we will be putting $\kappa_{np}=\kappa_{sp}=0$. The
ansatz is then $w(z)=w_0(z)$ with $\vq$ and $\lambda$ as the
variational parameters. The energy minimization procedure and results
will be described in the next section. Here we discuss some of the
important features of the spin and charge density distributions
corresponding to this ansatz.

Since there is exactly one skyrmion per unit cell, there is one north pole 
and one south pole per unit cell. From the form of 
$w_0(z)$ given in equation (\ref{w0}), it is obvious that the north poles 
are at $\{\vR\}$. As we will now show, for a given lattice, $\vq$ determines 
the positions of the south poles. First consider a system of two skyrmions 
located at $\vR$ and $-\vR$. This configuration is described by,
\bea
w(z) = \lambda(\frac{e^{i \vq.\vR}}{z-\vR} + \frac{e^{-i
\vq.\vR}}{z+\vR})
\eea
The zero of this ansatz is located at $-i \vR tan(\vq.\vR)$. Thus given the 
positions of the two skyrmions, the position of the zero is determined
by $\vq$. 

We will now illustrate that the position of the zero is 
determined by the value of $\vq$ by considering the case of a 
face-centered rectangular lattice. 
These lattices are described by the basis vectors $e1$ and $e2$. The unit cell
is also shown. The center of the unit cell is located at $z_0 = (e1+e2)/2$. 
Let us first consider N\'eel order $\vq = (\pi,\pi)$. Now,
\bea
w(z_0) = \Sigma_{nm} \frac{e^{i \pi (n +m)}}{(n-1/2)e1 + (m-1/2)e2}
\eea
Reflecting every lattice point about the point $z_0$ by the transformation $n = -n^{\prime}
-1$ and $m = m^{\prime}-1$, we can show that the above equation becomes
\bea
w(z_0 ) &=& \Sigma_{n^{\prime}m^{\prime}} \frac{e^{i \pi (n^{\prime} +m^{\prime})}}
{(-n^{\prime}+1/2)e1 + (-m^{\prime}+1/2)e2} \nn \\
& = & -w(z_0) 
\eea
which implies $w(z_0) =0$. Hence the point $z_0$ which is the center of the unit cell 
of the Bravais lattice is the location of the zero of the ansatz or the south pole.       
We note here that the point $z_0 = (e1+e2)/2$ is the south pole independent of 
specific values of $e1$ and $e2$ as long as there is N\'eel order. 
Hence in a triangular lattice and a rectangular lattice the south pole 
will lie at the center of the unit cell when these lattices are N\'eel ordered.

Now consider a triangular lattice with $\vq=(2\pi/3,2\pi/3)$ ordering. We
will now show that the position of the zero lies at the point $c_0 = (e1-e2)/3$. 
A rotation of a lattice point $\vR$ about this point through an angle of $2 \pi/3$ 
takes it to a point $\vR^{\prime}$ which also lies on the lattice. 
We know,
\bea
w(c_0) = \Sigma_{nm} \frac{e^{i 2\pi/3(n+m)}}{(c_0 - \vR)}
\eea
On multiplying $e^{i 2\pi/3}$ to the numerator and denominator of $w(c_0)$ it becomes
\bea
w(c_0) = e^{i 2\pi/3}\Sigma_{n^{\prime}m^{\prime}} 
\frac{e^{i 2\pi/3(n^{\prime}+m^{\prime})}}{(c_0 - \vR^{\prime})}
\eea
Since $\vR^{\prime}$ is also a Bravais lattice vector
\bea
w(c_0) = e^{i 2 \pi/3} w(c_0).
\eea 
Hence, we find that $c_0 = (e1 - e2)/3$ is the south pole of the ansatz 
for a triangular lattice with $(2 \pi/3,2 \pi/3)$ order. 
     
When the ordering on a triangular lattice changes from N\'eel
$(\pi,\pi)$ to $(2 \pi/3, 2 \pi /3)$, the position of the south pole
moves from $z_0 = (e1+e2)/2$ to the point $c_0 = (e1-e2)/3$. Hence we
see that the position of the south pole of the ansatz on a fixed
lattice is controlled by the parameter $\vq$.  In the rest of this
section we will consider a rectangular lattice with N\'eel order and
an aspect ratio of $a/b$. The south poles of the ansatz are located at
the centers of the unit cells.

We will now study the properties of the ansatz on a rectangular lattice with N\'eel
order.
We show that there exists a special value of $\lambda=\lambda_s$ at which the charge
and spin densities are symmetric about the north poles and south poles. This
is true if the ansatz obeys the condition
\bea
\label{condition}
w(z+(e1+e2)/2) = \frac{1}{w(z)} ~~~~ at ~~~\lambda = \lambda_s
\eea
 and hence the charge density and spin polarisation density are
\bea
\rho(w) &= \frac{1}{2 \pi i} \frac{\epsilon_{ij} \partial_i w \partial_j \wbar} 
{(1+w \wbar)^2} =& \rho(1/w) 
\eea
\bea
s_z(w) &= \frac{1}{2 \pi}\frac{1 - w\wbar}{1+w\wbar} = & - s_z(1/w) \nn
\eea 
Thus, if Eq.\ \ref{condition} is true, 
the charge density is symmetric near the north and south poles and the 
spin is also symmetric but in opposite directions near these poles. 

We impose the condition in Eq.\ (\ref{condition}) 
close to the north and south poles and obtain an equation
for $\lambda_s$ as follows. Expanding $w(z=z_{np}+ \epsilon)$ about the north pole 
($z_{np}$) we find that to
leading order $w(z) = \frac{\lambda}{\epsilon}$. 
Expanding about the position of the south pole $z_0 = (e1+e2)/2$ we get
\bea
w(z) &=& w(z_0) + \epsilon \partial_z w(z) \Big|_{z_0} \nn \\
&=& \epsilon \Sigma_{n_1,n_2} \frac {\lambda (-1)^{n_1+n_2}}{(z_0-(n_1 e1+n_2 e2))^2}.
\eea
The condition in Eq.\ \ref{condition} means that $\lambda_s$ obeys
\bea
\frac{1}{\lambda_s^2} =  \Sigma_{n_1,n_2} \frac{(-1)^{n_1+n_2}}{(z_0-(n_1 e1+n_2 e2))^2}.
\eea
This was solved for numerically for $\lambda_s$ and the condition in Eq.\
\ref{condition} was checked for all $z$. For example, for a N\'eel 
ordered rectangular lattice with an aspect ratio of $\sqrt{3}$, $\lambda_s$ was
numerically calculated to be equal to $8.42$. Even though the $\lambda_s$ was
obtained by imposing Eq.\ \ref{condition} at points close to the north and
south poles, the numerical calculation described above shows that 
Eq.\ \ref{condition} is satisfied at all $z$. Thus at $\lambda_s$ the charge
and spin densities are symmetric about the north and south poles.   

Having shown that at a particular value of $\lambda=\lambda_s$ the
charge and spin densities are symmetric around the north and south
poles of a rectangular lattice with N\'eel order, we motivate some
more properties of this configuration which will be numerically
verified in the next section. For such a rectangular lattice with an
aspect ratio of $\sqrt{3}$, the north poles and south poles form a
smaller triangular lattice. At the value $\lambda = \lambda_s$, the
charge and spin density also have the symmetry of this smaller
triangular lattice since they are symmetric about the north and south
poles (see fig.\ 3).  Since such a configuration corresponds to a
charge distribution with the triangular lattice symmetry we expect
that the Coulomb energy will be minimised for such a distribution.
For the special case of the square lattice with aspect ratio $1$, the
charge and spin densities also have the symmetry of a smaller square
lattice at the value $\lambda_s$.  This is our motivation for
minimising the variational ansatz for $w(z)$ with the variational
parameters as $\theta$, $q$, $\lambda$ and aspect ratio. We do this
calculation numerically to find the optimum value for these parameters
in the lowest energy configuration.

In order to contrast the behaviour of spin and charge density for
overlapping, large skyrmions with that of well separated, small
skyrmions we plot the XY spin configurations and charge density
profiles for a lattice of overlapping skyrmions. Figures 6. and 7.
s
how these plots for a rectangular lattice with an aspect ratio of
$\sqrt{3}$. We see that the XY spin configuration is very different
from that shown in Figure 2.  Also, the charge orders into a honeycomb
lattice which has the symmetry of a triangular lattice. Figures. 8.
and 9. show the plots for a square lattice of overlapping skyrmions.
    
We note here that the above features of the analytic ansatz remain true 
for a general ansatz $w(z) = w_0(z)F(z)$ as long as $F(z)$ is a function 
that respects the north pole-south pole symmetry. We discuss below a 
certain type of such functions and also conditions under which the north pole-
south pole symmetry is broken.

In the low and intermediate stiffness regimes the gradient term does
not dominate the energy functional anymore. Hence the requirement of
the analyticity of the ansatz is not valid any longer.  The ansatz can
have an exponential cut-off via the function $F(z)$.  In view of the
north pole-south pole symmetry (which is still valid at $g=0$), we
choose to damp the ansatz around the north and south pole in a
symmetric way.  Accordingly we choose, $\kappa_{np} = \kappa_{sp}$ in
Eq.\ (\ref {fz}).  This form for $F(z)$ preserves the north pole-south
pole symmetry.

Symmetry about the north and south poles can be broken in two ways. 
One way is by choosing the scale $\lambda$ to be different from the 
symmetric value $\lambda_s$ thus making the charge and spin densities 
asymmetric about the north and south poles. The other way is to use different 
exponential cut-off scales around both poles ({\it i.e } $\kappa_{np} \ne \kappa_{sp}$). 
This can be accomplished by
\bea
F(z) = \frac{e^{-\kappa_1 r_{np}}}{e^{-\kappa_2 r_{sp}}} 
\eea
Such an asymmetry can be induced by a large value of $g$. In the extreme 
case when the cut-offs are such that the charge is peaked entirely around 
the north pole, the overlap is very small and the oriented disc picture is valid.

\section{Energy minimisation : Ground state configurations}
In this section we show that the special value $\lambda_s$ at which
the north and south poles are symmetric is the one at which the energy
is minimised. We also show below that in the minimum energy
configuration the charge density has the symmetry of a triangular
lattice in the high stiffness limit and that of a square lattice in
the low stiffness limit.  This corresponds to the skyrmion lattice
being rectangular with an aspect ratio of $\sqrt{3}$ in the former
case and square in the later case.  In intermediate stiffness regimes,
the aspect ratio interpolates between $\sqrt{3}$ and $1$.  At the
minimum, the spin density is peaked equally in opposite directions at
the north and south poles which means $\langle s_z \rangle = 0$ in the
unit cell.  The vanishing of the $z$-component of the spin the unit
cell at $\lambda=\lambda_s$ is a universal feature of our
calculations.

\subsection{High Stiffness Regime}
In this regime at $g=0$, the NLSM energy functional has contributions
from the gradient term and the Coulomb term. Since the coefficient
multiplying the gradient term, $\gamma$ (the spin stiffness), is large
the gradient energy dominates the energy functional.  We minimize the
energy functional with the analytic ansatz defined in the previous
section. As mentioned earlier, this ansatz minimizes the gradient term
exactly and hence is a good choice in the high stiffness limit.  For
large, overlapping skyrmions, we expect that the charge density has a
higher symmetry than the underlying lattice.  Based on the arguments
in the previous section, we can see that this higher symmetry involves
the symmetric distribution of charge and spin around the north and
south poles.  The charge density has the symmetry of a face-centered
rectangular lattice if the underlying skyrmion lattice is rectangular.

We numerically minimize the NLSM energy functional with the ansatz for
$w(z)$ on a rectangular lattice with the quantities $\vq$,$\lambda$
and aspect ratio $a/b$ as the variational parameters.  N\'eel
orientation order ({\it i. e.} $\vq = (\pi,\pi)$) is chosen by the
minimum energy configuration. This ensures that the south pole lies at
the center of the rectangular unit cell. An aspect ratio of $a/b =
\sqrt{3}$ is chosen by the minimum energy configuration (see Fig.\ 
10).  At these values for $\vq$ , $a/b$ and $\lambda = \lambda_{min}$
the charge density is symmetrically distributed around the north and
south poles of the distribution as shown in Fig.\ 11.  Hence this
value of $\lambda = \lambda_{min}$ is also the special value
$\lambda_s$ discussed in the previous section at which there is
complete symmetry between the north and south poles. The charge
density has the symmetry of a triangular lattice at this point.

The spin density peaks equally in opposite directions at the north
pole and south pole. Hence the most symmetric distribution corresponds
to a situation in which the total $z$-component of the spin is zero.

Because of the absence of the competing Zeeman term, there is no other
length scale competing with the length scale of the Coulomb energy.
The scaling relation is $E_{coul}(\lambda,\vq) = \frac{1}{\lambda}
E_{coul}(1,\vq)$ (see Appendix 1).  As the filling factor is varied,
the length scale corresponding to the Coulomb energy changes.  Thus,
the length scale $\lambda$ should scale with the filling factor
(lattice spacing) but the ground state configuration does not change.
If the energy is minimised at a certain filling factor (corresponding
to lattice spacing $e_1$) by a value $\lambda_1$, then the energy at
another filling factor (corresponding to another lattice spacing
$e_2$) is minimised by a value $\lambda_2$ such that $\lambda_1/e_1 =
\lambda_2 /e_2$ is satisfied. This scaling of $\lambda_{min}$ with the
lattice parameter $e$ is shown as $\nu$ is changed (see Fig.\ 12).
The ground state remains a rectangular lattice with aspect ratio of
$\sqrt{3}$ with N\'eel order for all filling factors at $g=0$.

To conclude this subsection we give some details regarding the actual
numerical minimization procedure used and some checks that were done.
The numerical routine used was a downhill simplex routine in the three
parameter space of $\vq$,$\lambda$ and $a/b$. This routine minimized
the energy functional with respect to the above parameters and
returned the minimised value of energy at the appropriate values of
these parameters. The minimization was carried out for many filling
factors close to $1$. As explained in Section II, different filling
factors correspond to varying the area of the lattice. The numerical
code was first checked to give the correct topological charge of $1$
per unit cell (with an error of less that $0.1\%$).  The vanishing of
the $z$-component of the spin was numerically seen when
$\lambda=\lambda_{min}$, lending support for our expectation of
complete north pole-south pole symmetry at this point.

\subsection{Low and Intermediate Stiffness Regimes}
We now consider the opposite limit (zero stiffness limit) at $g=0$ for
a system of overlapping skyrmions. In this limit only the Coulomb
energy contributes to the energy functional at $g=0$ and hence minimum
energy configurations of the skyrmion lattice are determined by the
minimum of the Coulomb energy.  The Coulomb energy is minimized by a
suitable configuration in which the charge is uniformly distributed.
The ansatz used for the minimisation need not be analytic any longer
and can have an exponential cut-off. Keeping in mind the north and
south pole symmetry we chose to damp the ansatz around the north and
south poles in a symmetric way. This is done by choosing $F(z)$ to be
a function with the periodicity of the north pole-south pole structure
in the unit cell. The form for $F(z)$ is given in Eq.\ (\ref{fz}) with
$\kappa_{np} = \kappa_{sp} =\kappa$.

The energy is minimized with this ansatz on a Bravais lattice with
respect to $\vq$, $\lambda$, $\kappa$ and aspect ratio $a/b$.  We find
that the ground state configuration is a square lattice with N\'eel
order and $s_z=0$ (see Fig.\ 13). The type of lattice does not change
with filling factor and the scaling of $\lambda$ and Coulomb energy
with lattice size is observed.  The charge density has the symmetry of
a square lattice.  The spin density is also peaked in opposite
directions at the north and south poles.

We have also checked whether an ansatz which describes
a charge density with a different behaviour at small $r$ will change the
features described above. For this purpose we chose an ansatz
\bea
w(z) &=& w_0(z) F(z) \nn \\
F(z) &=& \frac{e^{-\kappa_1 r_{np}}}{e^{-\kappa_1 r_{sp}}}
\frac{(1+\kappa_2 r_{np})}{(1 + \kappa_2 r_{sp})}
\eea
ensuring the symmetry of the north and south poles.
When the minimization was done for this ansatz we find that N\'eel order
and $s_z=0$ are still preferred with a square lattice as the ground
state at all filling factors. 
At filling factors below $\nu = 0.93$, we find that $\kappa_2 <<
\kappa_1$
and the charge density has the same behaviour as shown in Fig.\ 13.
Between filling factors $0.93$ and $0.98$ we find that
$\kappa_2>\kappa_1$  and the
 charge gets distributed to the lattice points as well 
as the center of the unit cell.

In the intermediate regime the stiffness is chosen to be at its
physical value $\gamma$. All three terms (gradient energy, Zeeman
energy and Coulomb energy) contribute to the energy functional.  The
ansatz used in Eq.\ (\ref{latans}) is used to determine the ground
state configuration at $g=0$. We find that the minimum energy
configuration is a rectangular lattice with an aspect ratio which is
ansatz and filling factor dependent.  However the spin configuration
still has N\'eel order and $s_z=0$ for all these configurations.

\section{Experimental consequences}

Our calculations predict two universal features of the $g=0$ system.
(i) The skyrmion lattice is rectangular with N\'eel orientational
order, (ii) the spin polarization is exactly zero.  It is difficult to
directly experimentally observe skyrmion lattices and no such
experiments have been done so far. However the spin polarization can
and has been measured using NMR and optical techniques at $g=g_0$
\cite{barrett,aifer,kukushkin}.

Our calculations are done at zero temperature and in absence of
disorder. However, disorder is always present in real systems and is crucial
to the physics of plateau formation in quantum Hall systems. Thermal 
effects are also expected to be important at the temperatures ($\sim 1K$)
where the experiments are typically done. In fact theoretical \cite{green,timm}
and experimental \cite{bayot} evidence indicates that the skyrmion
lattice would have melted at these temperatures. It is therefore important
to consider the effects of thermal fluctuations and disorder before 
making predictions for experiments.

\subsection{Spin Polarization at $g=0$}

In this section, we present a general proof for the vanishing of the 
thermal average of the spin polarization at $g=0$ in the presence of 
an arbitrary static disorder potential, $V_{dis}(x,y)$. 
The classical partition function can be written as,
\bea
Z  &=& \int_{w,\wbar} e^{\beta E[w,\wbar]+\beta \int d^2x
V_{dis}(x,y) \rho(x,y)} 
\eea 
The spin polarization of the system (of size L) is given by the thermal average
\bea
\label{thav}
\langle SP \rangle &=& \frac {1}{\pi L^2} \frac{1}{4 \pi}
\frac {1}{Z}
 \int_{w,\wbar} \frac{1-w \wbar}{1+w \wbar}~~
e^{\beta E[w,\wbar]+\beta \int d^2x 
V_{dis}(x,y) \rho(x,y)} 
\eea
Note that at $g=0$, the energy functional depends only on the charge density
and is $O(3)$ invariant.
If we make the transformation of $w^{\prime}\rightarrow = 1/w$, 
the charge density and hence the energy density are
invariant but the spin density changes sign.
Equation (\ref{thav}) then shows that the average spin polarization vanishes. 
However, this argument for proving that spin polarization vanishes on average
is technically incorrect since the system is realised in a broken symmetry 
phase. Thus the average in equation (\ref{thav}) is only over configurations 
satisfying the boundary conditions,
\be
\label{bc}
\lim_{\vert z\vert \rightarrow \infty}w(z,\bar z) = w_\infty
\ee
where $w_\infty$ is a fixed value. Since $w^{\prime}$ and $w$ clearly cannot
both satisfy the above condition, the above argument is flawed.

We now give a more sophisticated argument which takes into account the above 
point and prove the result for a large class of configurations. Without loss 
of generality, we can put $w_\infty = 0$.
We first consider analytic configurations of an an $N$ skyrmion system,
\bea
w(z) = \sum_{n=1}^N \frac{\lambda_n}{(z-\vR_n^{np})}
\eea
Where $\vert \vR_n^{np}\vert \le L$ and the labelling is such that 
$n > m \Rightarrow |R_n^{np}|>|R_m^{np}|$. Let $\vR_n^{sp}$ denote the 
zeroes of $w$. The boundary condition, $w_\infty=0$, ensures that atleast
one of the zeroes lies at infinity. $w$ can then be written as a product of 
the positions of the poles and zeros:
\be
w(z)=\lambda \frac{\prod_{n=1}^{N-1} (z-\vR_n^{sp})}
{\prod_{n=1}^N(z-\vR_n^{np})} \nn \\
\ee
where,
\be
\lambda = \sum_{n=0}^{N}\lambda_{n}
\ee
If $\lambda = 0$ implies that there is more than one zero at infinity.
For every such configuration, there are a very large number of nearby
configurations of the type $\lambda_n = \lambda_n + \delta\lambda_n$, 
for which $\lambda\ne 0$. Thus the set of $\lambda=0$ configurations has
negligible statistical weight. So for a generic analytic configuration,
only one zero lies at infinity. We concentrate on these configurations 
and rewrite $w$ as, 
\be
w(z)= \frac {\lambda}{z - \vR_N^{np}} \prod_{n=1}^{N-1} 
\frac {z-\vR_n^{sp}}{z-\vR_n^{np}} \nn\\
\ee  
Now consider the configuration,
\be
w^{\prime}(z)=\frac{1}{\lambda} \frac {(\vR_N^{np})^2}{(z-\vR_N^{np})}
\prod_{n=1}^{N-1} \frac {z-\vR_n^{np}}{z-\vR_n^{sp}} \nn \\
\ee
$w^\prime$ satisfies the boundary conditions (\ref{bc}). In the region 
$|z| << {\vR_N^{np}}$, we have
\bea
\label{bulkw}
w^{\prime}(z) &\approx& -\frac{\vR_N^{np}}{\lambda} 
\prod_{n=1}^{N-1} \frac {z-\vR_n^{np}}{z-\vR_n^{sp}} \nn \\
&\approx& \frac{1}{w(z)}
\eea
Thus, for every configuration $w(z)$, there exists another configuration
$w^{\prime}(z)$ which satisfies the same boundary conditions and except in 
a region around one point near the edge, is equal to $1/w(z)$. The important 
fact is that the area of this region does not change as we take the 
thermodynamic limit, $N,L\rightarrow \infty,~N/(\pi L^2)\rightarrow \rho$. 
For short range interactions we then have,
\bea
\label{bulkesp}
E[w]&=&E[w^\prime] + o(L^0)\nn \\
SP[w]&=&-SP[w^\prime] + o(L^0) 
\eea
Note that a $1/r$ interaction in two dimensions is short-ranged in this sense.
Equation (\ref{thav}) then shows that the average spin polarization
density vanishes in the thermodynamic limit.

We now generalize this proof for arbitrary configurations that have only one
zero at infinity. Consider configurations that satisfy,
\bea
\lim_{\vert z\vert\rightarrow\infty}w(z,\bar z)&=&0 \nn \\
\lim_{\vert z\vert\rightarrow\infty}zw(z,\bar z)&=&\lambda
\eea
Define,
\be
w^\prime(z,\bar z)=\left(\frac{R_N^{np}}{z-R_N^{np}}\right)^2
\frac{1}{w(z,\bar z)}
\ee
It can easily be verified that $w^\prime$ satisfies the boundary conditions
(\ref{bc}) and that equations (\ref{bulkw}) and (\ref{bulkesp}) hold true.
Thus if we assume that, as has been shown for analytic configurations, the
generic configuration has only one zero at infinity, we have proved that the
spin polarization density will vanish in the thermodynamic limit.

\subsection{Spin Polarization at $g\approx 0$}

The results obtained in the previous section show that for a system 
with $g$ exactly zero, the spin polarization  
changes discontinuously from its fully polarized value at $\nu=1$ to
zero when the filling factor changes from 1. In a real system, $g$
will never be exactly equal to zero but will have some very small
value. Correspondingly, the single skyrmion at $\nu=1$ will have a
large but finite size. The $g\sim 0$ regime discussed in this paper
will occur at filling factors where this size becomes comparable to
the inter-skyrmion spacing, eg. in the low $g$ experiments
\cite{shukla,nick}, the single skyrmion spin at $\nu=1$ has been
estimated to be around 30-40. As mentioned earlier, for "hedgehog"
skyrmions, the spin, $s$ can be related to the rms charge radius as
$<r^2> = 2s$ (see Appendix 2). Setting the inter-skyrmion spacing,
$a=\sqrt{2\pi/(1-\nu)}$ equal to the charge radius, $\sqrt{2s}$ yields
$\nu \sim 0.9$. Thus $\nu \le 0.9$ will be the $g \approx 0$ regime
for these skyrmions.

We have calculated the spin polarisation at $g=0.05g_0$ in both the
high and low stiffness limits. The skyrmion lattice configuration that
minimizes the energy now has a value of $\lambda$ slightly away from
the symmetric point resulting in a non-zero spin polarisation. The
spin polarisation is calculated from 
\bea 
SP_{lattice} = \frac{\nu}{4
  \pi A_{\Box}} \int_{\Box} \frac{1-\overline{w} w}{1 + \overline{w}
  w} 
\eea 
The results are plotted in Figs.\ 14 and 15.  As can be seen
they are not very different. The spin polarization falls sharply to
about 10\% of its $\nu=1$ value at $\nu=.97$ and $\nu=.98$ in the high
and low stiffness limits respectively. Both are down to about 1\% at
$\nu=0.9$.  At this value of $g$, we have calculated the single
skyrmion spin, as described in Appendix 1 to be 45 and 33 in the low
and high stiffness limits respectively.  This implies that the spin
polarization has fallen to 10\% when the charge radius is equal to
half the inter-skyrmion spacing. i.e. when they are just touching each
other. It has fallen to about 1\% when the radius is roughly equal to
the lattice spacing. This is consistent with our general arguments of
the previous paragraph.

To summarize, we predict that the spin polarization will fall sharply
to about 10\% of the fully polarized value at $\nu=1-(\pi/4s)$, where
$s$ is the single skyrmion spin measured at $\nu=1$. It will fall to
almost zero at around $\nu=1-(\pi/s)$ and remain zero as $\nu$
decreases further. 

These predictions are based on our calculations done 
at zero temperature and in absence of disorder.  
However previous work indicates that the spin polarization
may be fairly insensitive to these effects. The Hartree-Fock
calculation of Brey et. al.  \cite{brey} was done for the pure system
at zero temperature. Nevertheless, the calculated spin polarization 
(at $g=g_0$) matched fairly well with the experiments of Barret et. al.
\cite{barrett} which were done at 1.55 K. Thus the predictions in the 
previous paragraph may well hold for real systems.

\section{Conclusions}
In this paper we have studied the ground state configuration of a
lattice of overlapping skyrmions in various stiffness regimes. Their
behaviour is qualitatively different from small, well-separated
skyrmions. While well-separated skyrmions can be described well by the
oriented, charged disc picture, for overlapping large skyrmions this
picture breaks down. A lattice of large skyrmions is described by the
form of the ansatz given in Section III which is specified completely
in terms of the positions and strengths of the poles and zeros.  The
north poles are the lattice points and the south poles determined by
the parameter $\vq$. In the case of N\'eel ordered, face- centered
rectangular lattices, the south poles lies at the center of the unit
cell. We then argued that there exists a special value of
$\lambda=\lambda_s$ when the charge and spin densities are
symmetrically distributed around the north and south poles. In the
case when the lattice was rectangular the charge density formed a
smaller face-centered rectangular lattice. When the lattice was square
the charge density also had the symmetry of a smaller square. In all
cases the spin density was peaked in opposite directions at the north
and south poles and the $z$ component of the spin vanishes. Since this
distribution corresponds to the most symmetric distribution of spin
and charge we expect that this also minimises the energy.

We verify the above arguments numerically in Section IV. We find that
in the high stiffness regime the underlying skyrmion lattice is N\'eel
ordered and rectangular (with aspect ratio $\sqrt{3})$ ) corresponding
to the charge density having the symmetry of a triangular lattice. The
value $\lambda_{min}$ which minimises the energy was found to be
exactly equal to the value $\lambda_s$ at which there was complete
symmetry between the north and south poles.  In the low stiffness
regime the underlying lattice and charge density had square symmetry.
In the intermediate regimes the lattice was rectangular with an aspect
ratio between $\sqrt{3}$ and $1$. In all cases the $z$-component of
the spin vanished.

We give a general argument for the vanishing for
the spin polarization at $g=0$ at finite temperature and in the
presence of disorder.
We also discuss some experimental consequences. We calculate the spin
polarisation at small but non-zero $g$.  We predict that this quantity
falls sharply to $10 \%$ of the fully polarised value at $\nu = 1-
\pi/4s$ and to almost zero at $\nu = 1- \pi/s$.

We conclude that in the regime of small Zeeman coupling, skyrmions are
large and overlapping and the oriented, charged-disc picture breaks
down.  This is because there is a distribution of charge and spin
around the south poles also unlike in the charged-disc picture where
they were distributed only around the north poles. The charge and spin
densities can have a higher symmetry than the underlying lattice in
the overlapping case unlike in the well-separated case. The universal
features of the classical ground state are (1) N\'eel ordered
rectangular lattices of skyrmions with the charge density having the
symmetry of a face-centered rectangular lattice and (2) vanishing of
the $z$-component of the spin.

\section*{Appendix 1}
In this appendix we calculate the scaling form of the single skyrmion
energies with the parameter $\lambda$ and also the spin polarisation
of a lattice of such skyrmions. 
The ansatz chosen to minimize the energy of a single skyrmion with winding number $1$ at $g=0$ is of the form
\bea
w(z) = \frac{\lambda}{r}e^{-\kappa r}
\eea
The ansatz is not analytic anymore and has an exponential damping factor. From the form
of the individual energies given in the previous sections we can see that they scale
with $\lambda$ as
\bea
E_{grn}(\kappa,\lambda) &=& E_{grn}(\kappa,1) \nn \\ 
E_{z}(\kappa,\lambda) &=& \lambda^2 E_{z}(\kappa,1) \nn \\ 
E_{coul}(\kappa,\lambda) &=& \frac{1}{\lambda} E_{coul}(\kappa,1)  
\eea
From this scaling relation we can calculate the $\lambda$ which minimizes the energy
\bea
\lambda_{min}^3 = \frac{E_c(\kappa,1)}{2 E_z(\kappa,1)}
\eea
Therefore the minimum energy is
\bea
E(\kappa,\lambda_{min}) = E_{grn}(\kappa,1)+\lambda_{min}^2 E_z(\kappa,1)+\frac{1}{\lambda_{min}}
E_{coul}(\kappa,1) 
\eea
The excess spin per skyrmion corresponding to the minimum configuration is 
\bea 
spin = \lambda_{min}^2 \frac{1}{2 \pi} \int \frac{\overline{w}w}{1 + \overline{w} w}
\eea
To extract the spin polarisation of a collection of well
separated $N_{sky}$ number of single skyrmions we have to sum over all the individual spins and 
subtract out the spin of the vacuum state ($spin_{vac})$. Thus, 
\bea
spin_{DL} &=& N_{sky}~spin - spin_{vac} \nn \\
\frac {spin_{DL}}{A} &=& \frac{spin}{A_{\Box}} - \frac{\nu}{4 \pi} \nn \\
&=& \frac{1}{2 \pi}((1-\nu) spin - \frac{\nu}{2})
\eea
where $A$ is the area of the sample and $A_{\Box}$ is the area of the unit cell of the dilute lattice formed by the
skyrmions. 

\section*{Appendix 2}
In this appendix we show the relation between the spin $s$ of a hedgehog skyrmion and 
its rms charge radius $r$ i.e. $\langle r^2 \rangle = 2 s$. Consider a single
skyrmion described by the ansatz 
\bea
w(z) = \frac{F(r)}{z}
\eea
where $r$ is the rms charge radius and $F(r)$ is an analytic function of the rms
charge radius. The spin of this skyrmion is
\bea
s = \frac{1}{2 \pi} \int \frac{F^2 d^2r}{r^2 + F^2}
\eea    
Its charge density is
\bea
\rho(r) &=& \frac{1}{ 2 \pi i} \frac {\epsilon_{ij} \partial_i \wbar \partial_j w}
{(1 + \wbar w)^2} \nn \\
& = & \frac{1}{ 2 \pi i} \partial_i \frac {\epsilon_{ij}  \wbar \partial_j w}
{(1 + \wbar w)}
\eea
Now,
\bea
\langle r^2 \rangle &=& \int r^2 \rho(r) d^2 r \nn \\
& = & \frac{\epsilon_{ij}}{ 2 \pi i}  \int r^2 \partial_i \frac { \wbar \partial_j w}
{(1 + \wbar w)} \nn \\
& = & -\frac{\epsilon_{ij}}{ 2 \pi i} \int \frac {2 r_i  \wbar \partial_j w}
{(1 + \wbar w)}
\eea
Using,
\bea
\partial_j w &=& \frac{\Fp r_j}{z r}+ F \partial_j (1/z) \nn \\
i.e~~ \partial_x w &=& -\frac {w}{z}~~~  \&\nn \\
\partial_y w &=& -\frac{ i w}{z} 
\eea
we get
\bea
\langle r^2 \rangle = 2 s
\eea

\newpage
\begin{figure}
{\epsfysize 2.7in \epsfbox{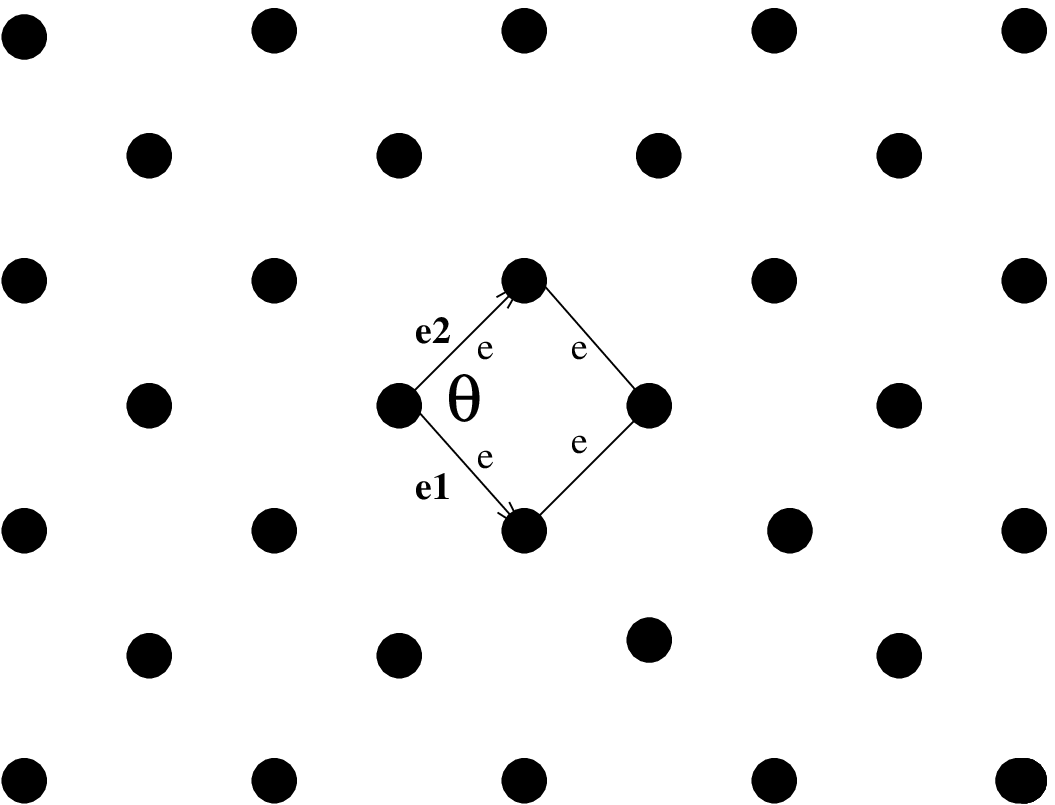}}\\{Fig.\ 1. 
The face-centered rectangular lattice with basis vectors}
\end{figure}

\begin{figure}
{\epsfysize 2.7in\epsfbox{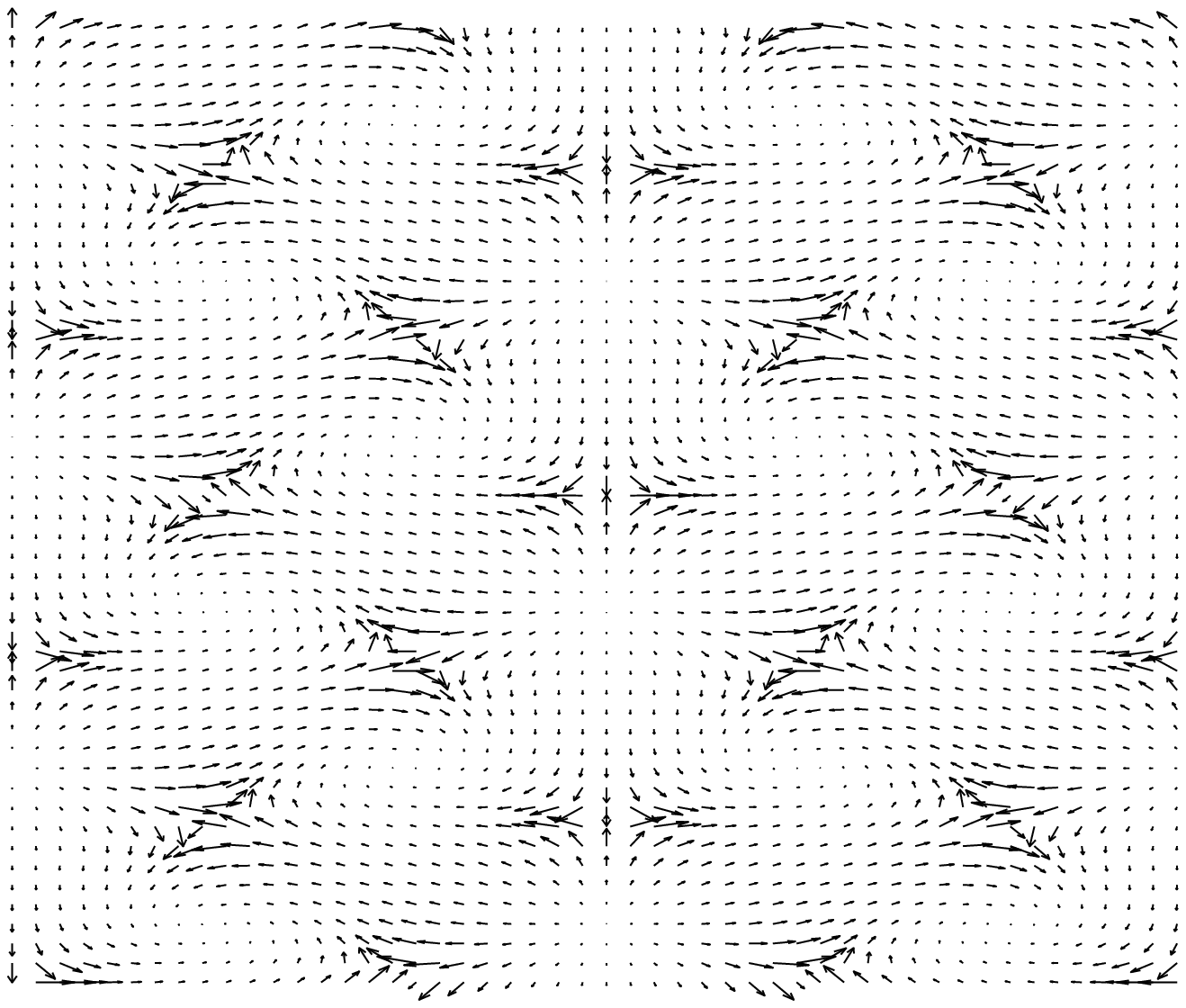}}\\{Fig.\ 2. XY spin configuration 
in a triangular lattice of well separated skyrmions ($\vq = (2 \pi /3,2 
\pi /3)$).
}
\end{figure}

\begin{figure}
{\epsfysize 2.7in\epsfbox{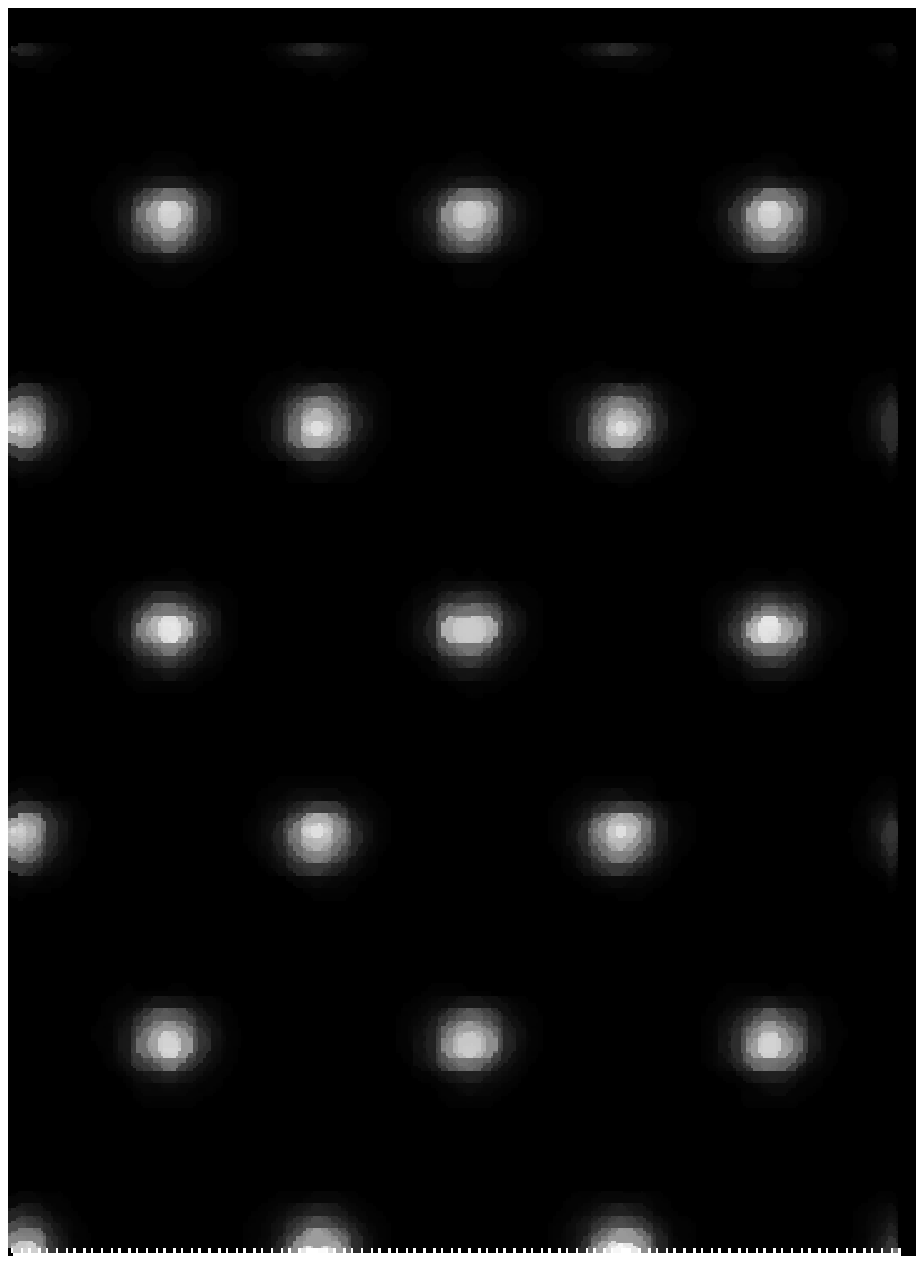}}\\{Fig.\ 3. Charge density profile 
in a triangular lattice of well separated skyrmions. The higher values
are shaded in white. }
\end{figure}

\begin{figure}
{\epsfysize 2.7in\epsfbox{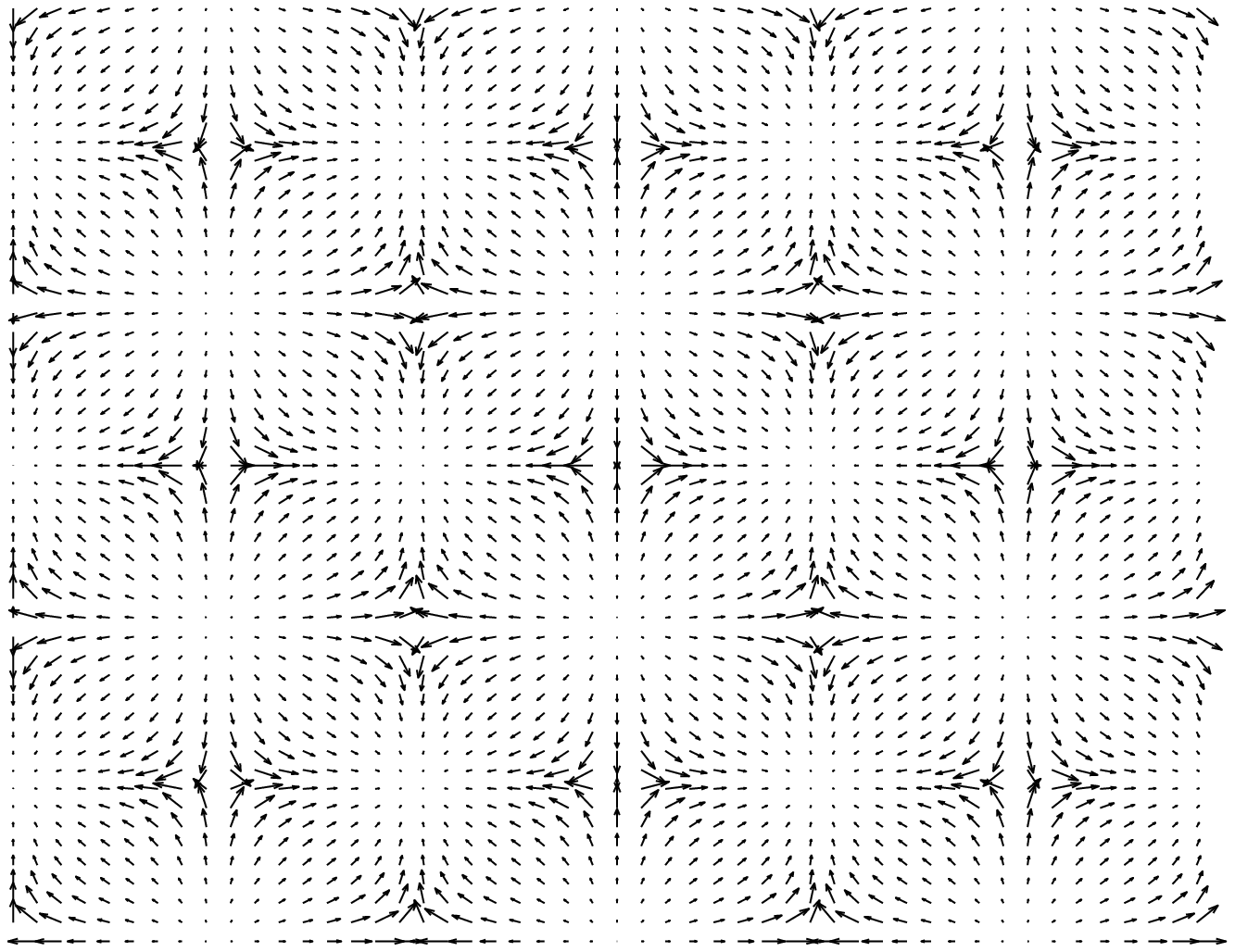}}\\{Fig.\ 4. XY spin configuration 
in a square lattice of well separated skyrmions ($\vq = (\pi,\pi)$).
}
\end{figure}

\begin{figure}
{\epsfysize 2.7in\epsfbox{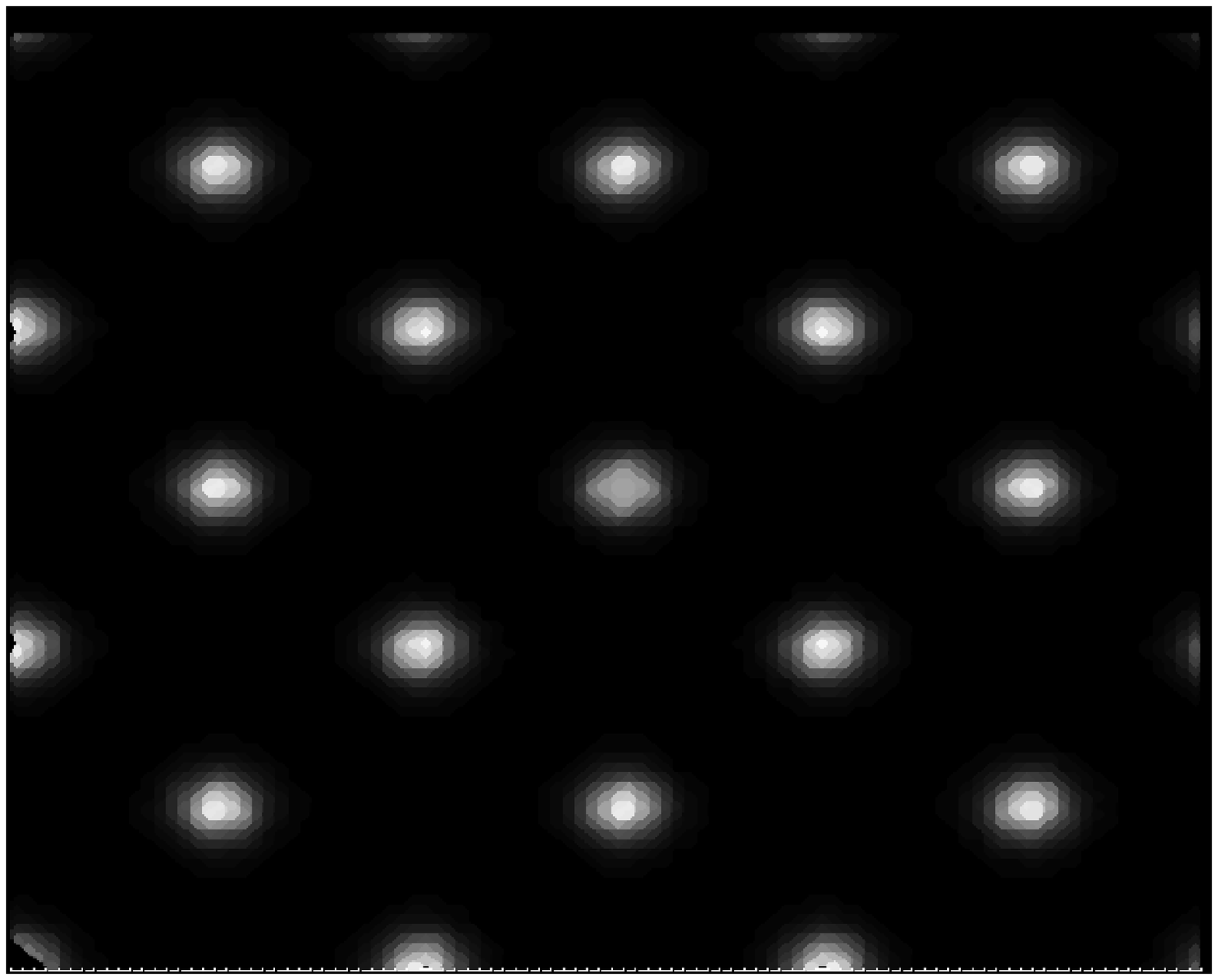}}\\{Fig.\ 5. Charge density profile 
in a square lattice of well separated skyrmions. The higher values
are shaded in white. }
\end{figure}

\begin{figure}
{\epsfysize 2.7in\epsfbox{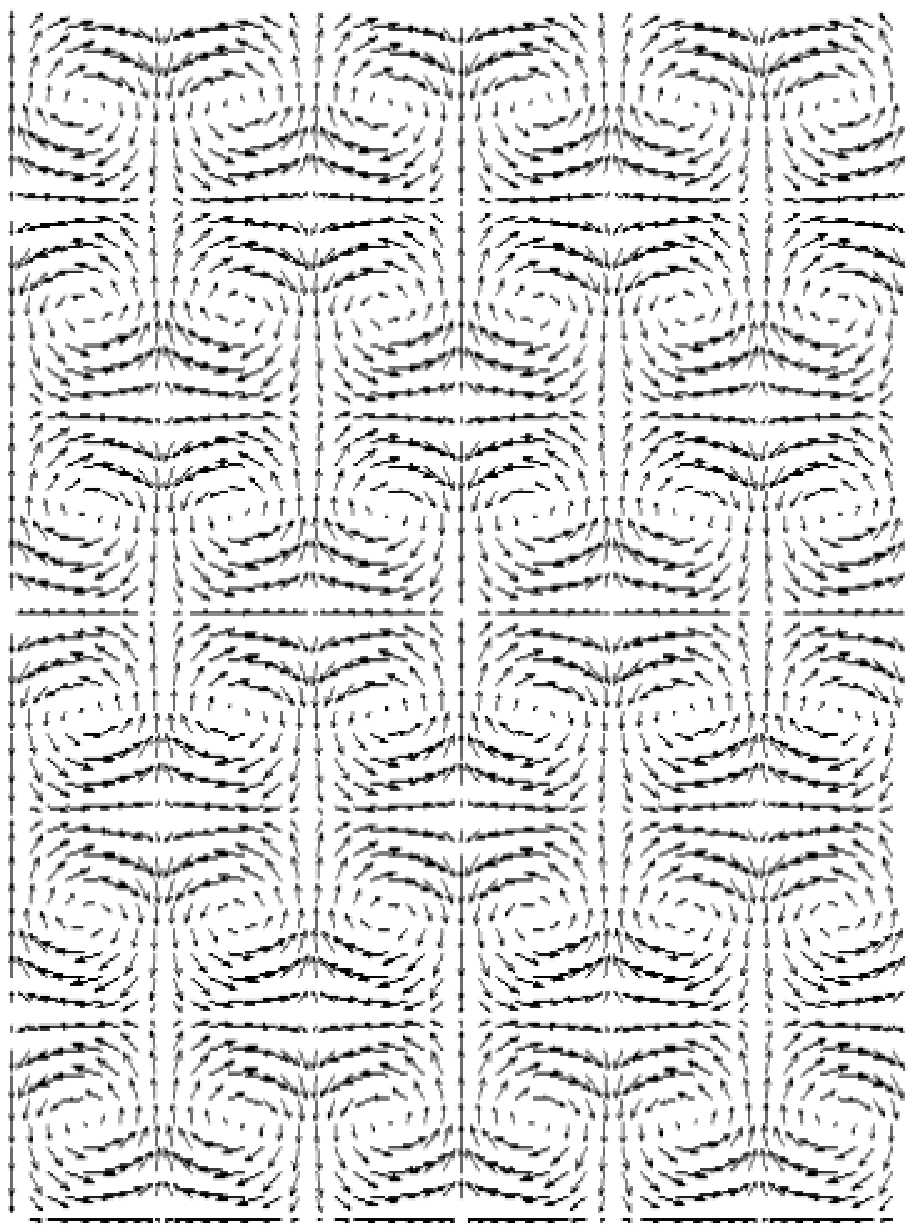}}\\{Fig.\ 6. XY spin configuration 
in a N\'eel ordered, rectangular lattice of overlapping skyrmions.
}
\end{figure}

\begin{figure}
{\epsfysize 2.7in\epsfbox{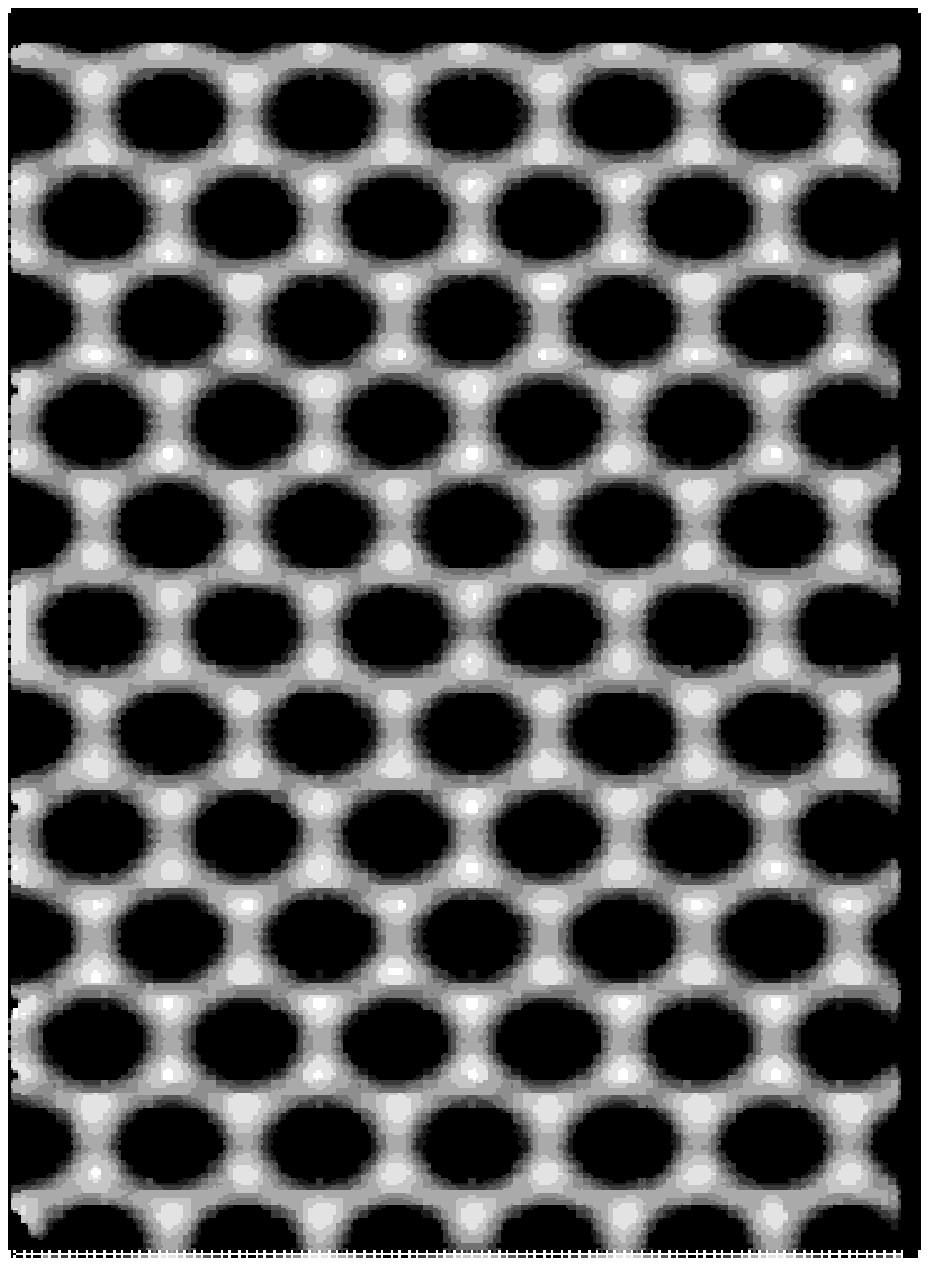}}\\{Fig.\ 7. Charge density profile 
in a N\'eel ordered, rectangular lattice of overlapping skyrmions. The higher values
are shaded in white. }
\end{figure}

\begin{figure}
{\epsfysize 2.7in\epsfbox{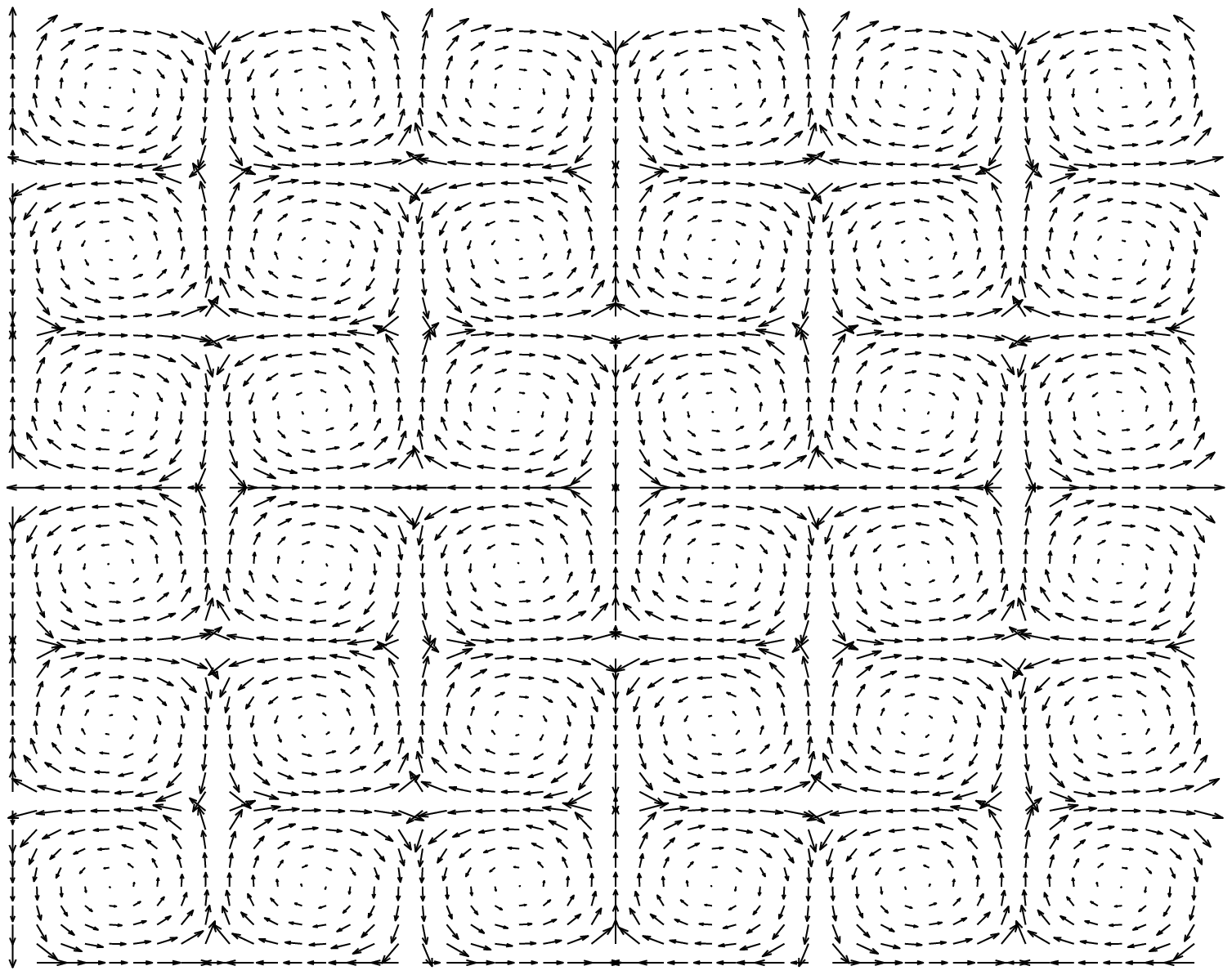}}\\{Fig.\ 8. XY spin configuration 
in a N\'eel ordered, square lattice of overlapping skyrmions.
}
\end{figure}

\begin{figure}
{\epsfysize 2.7in\epsfbox{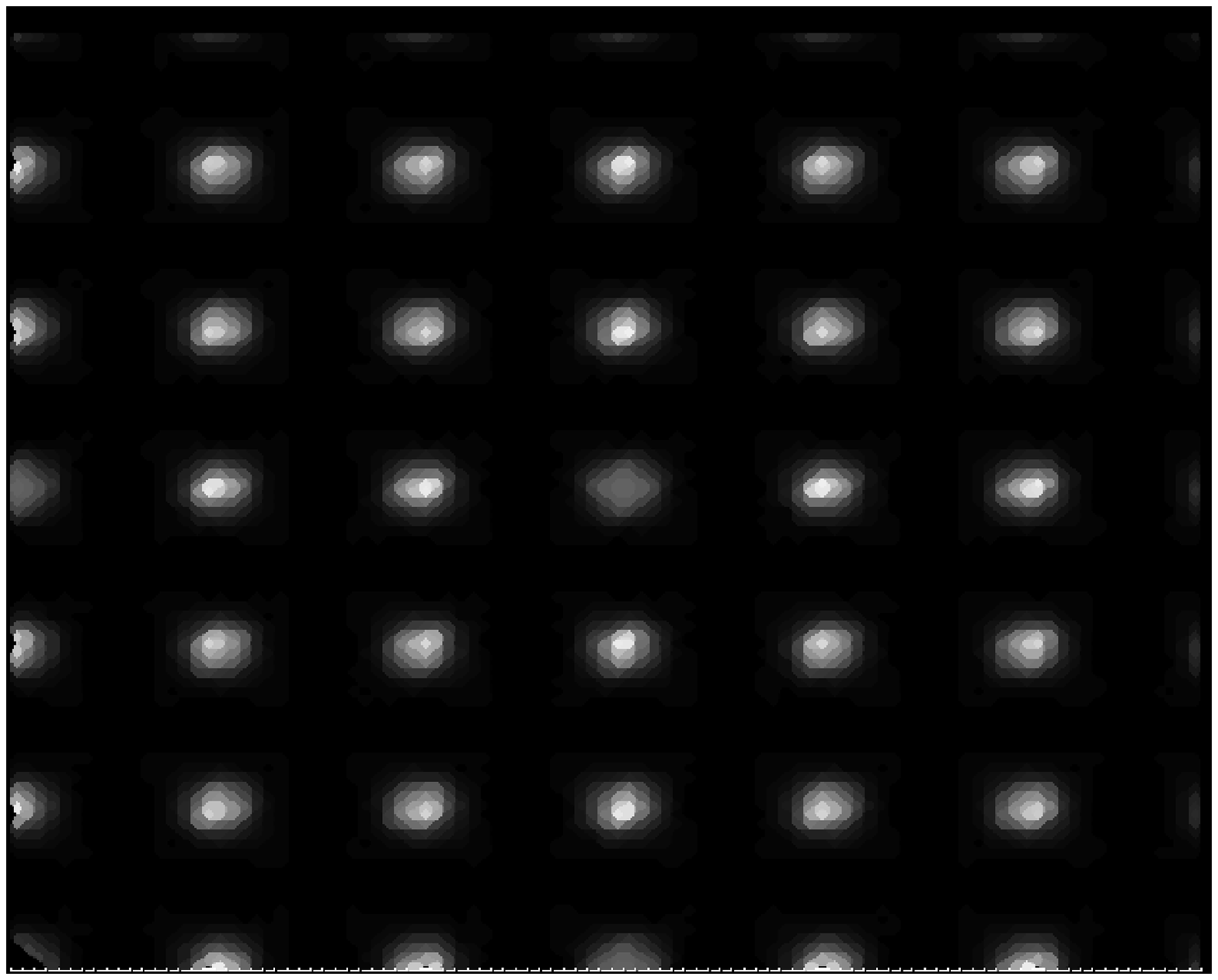}}\\{Fig.\ 9. Charge density profile 
in a N\'eel ordered, square lattice of overlapping skyrmions. The higher values
are shaded in white. }
\end{figure}

\begin{figure}
{\epsfysize 2.7in\epsfbox{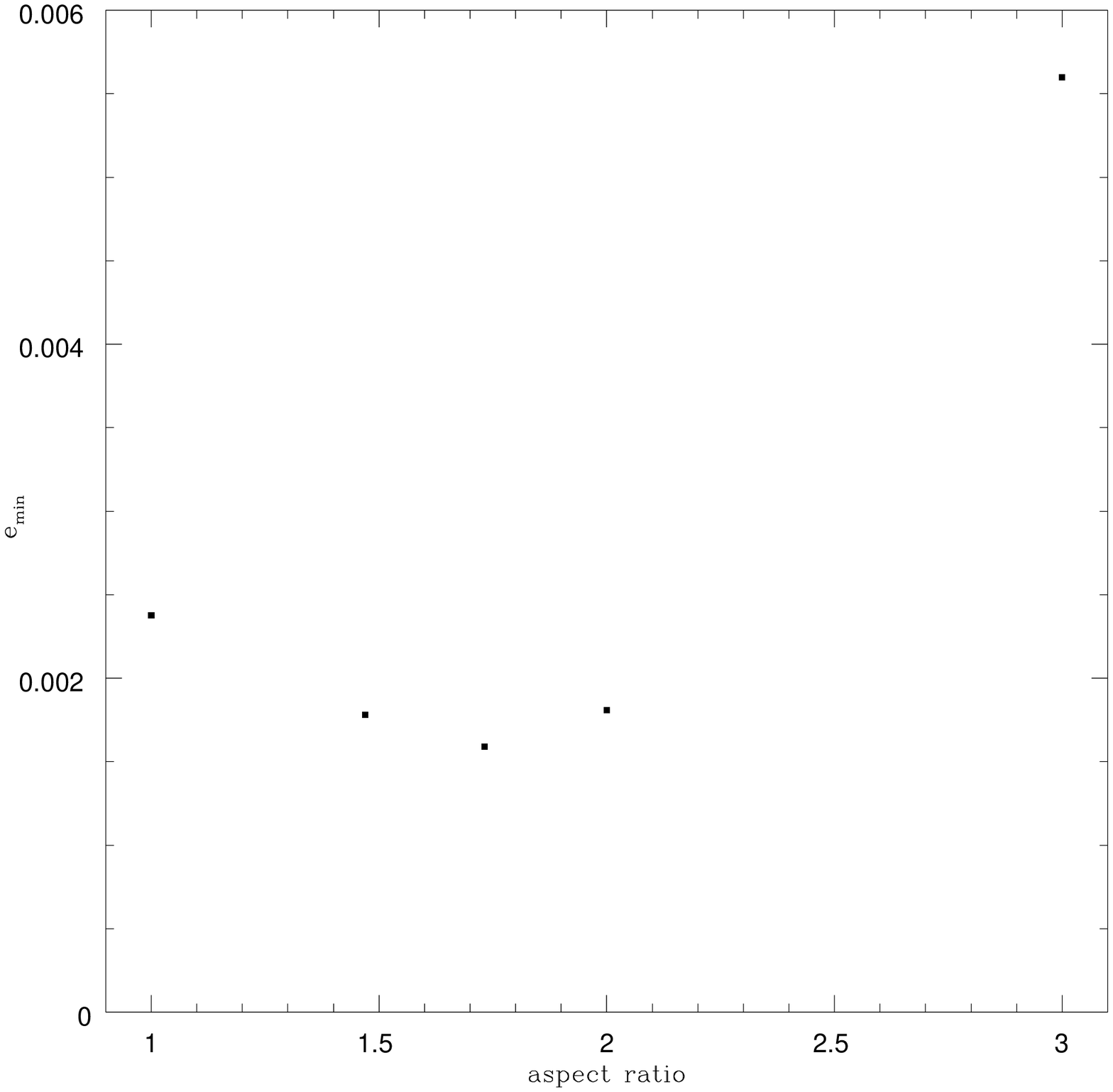}}\\{Fig.\ 10. 
Plot of the value of minimzed energy (with respect to $\lambda$ and $q$).
Notice that the minimum occurs when the aspect ratio
has value $\sqrt{3}$.} 
\end{figure}

\begin{figure}
{\epsfysize 2.7in\epsfbox{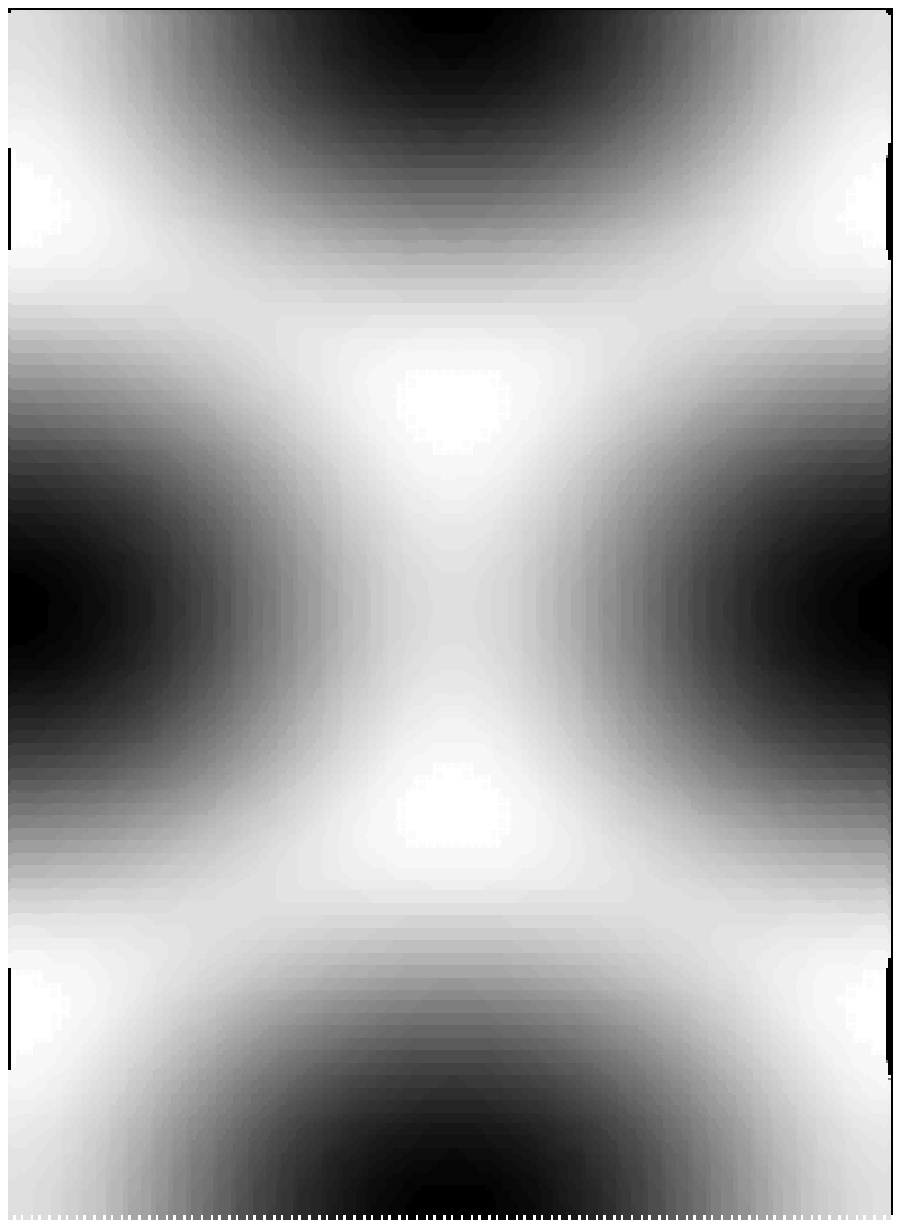}}\\{Fig.\ 11. Charge density profile in the unit cell of
the rectangular lattice (aspect ratio $\sqrt{3}$) at $\nu = 0.90$, in the high stiffness limit.
}
\end{figure}

\begin{figure}
{\epsfysize2.7in\epsfbox{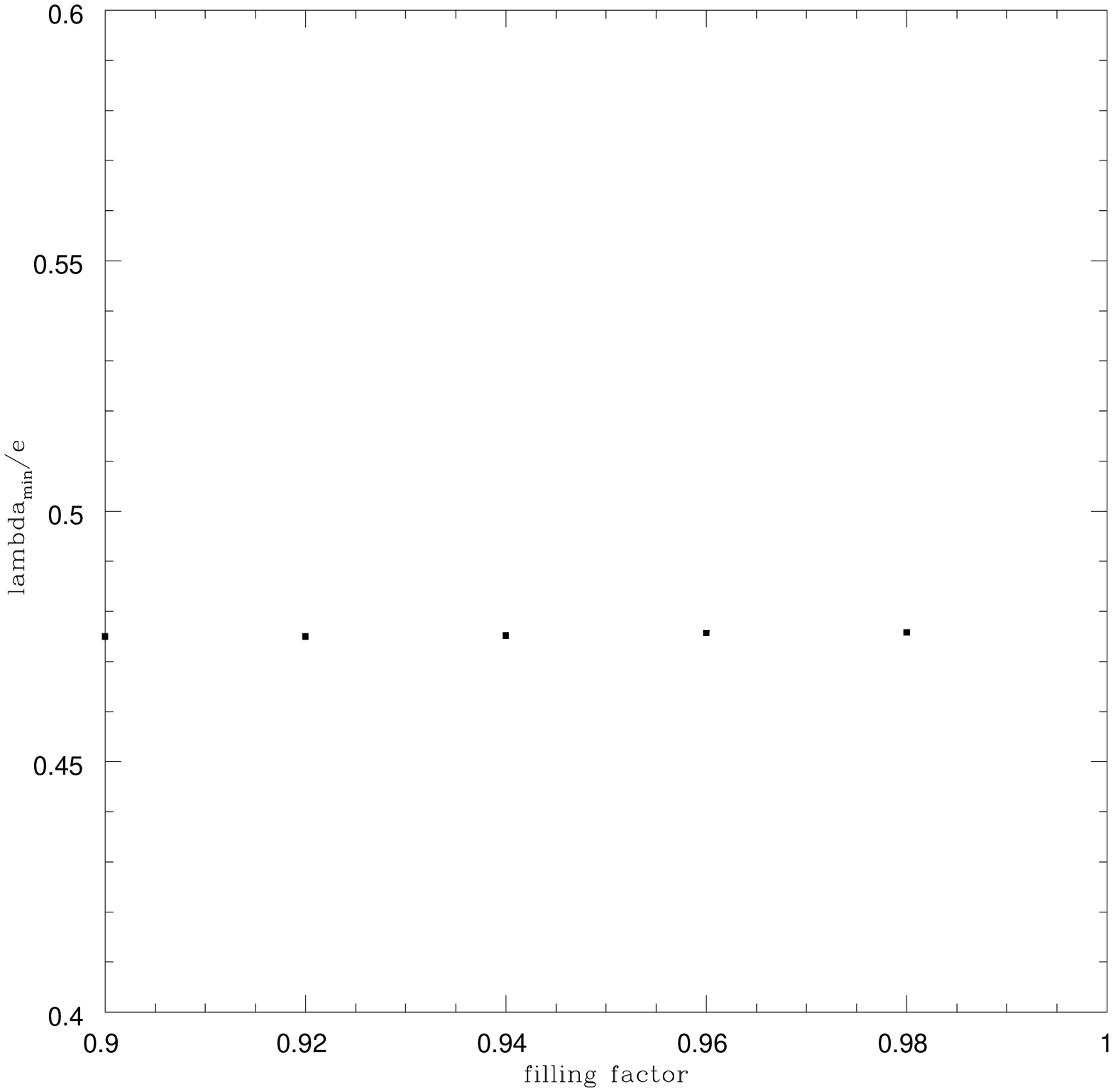}}\\{Fig.\ 12. Scaling of $\lambda$ with the lattice size paramter $e$ as a
function of the
filling factor in the high stiffness regime.
A similar graph is obtained in the low stiffness
regime also.}
\end{figure}

\begin{figure}
{\epsfysize 2.7in\epsfbox{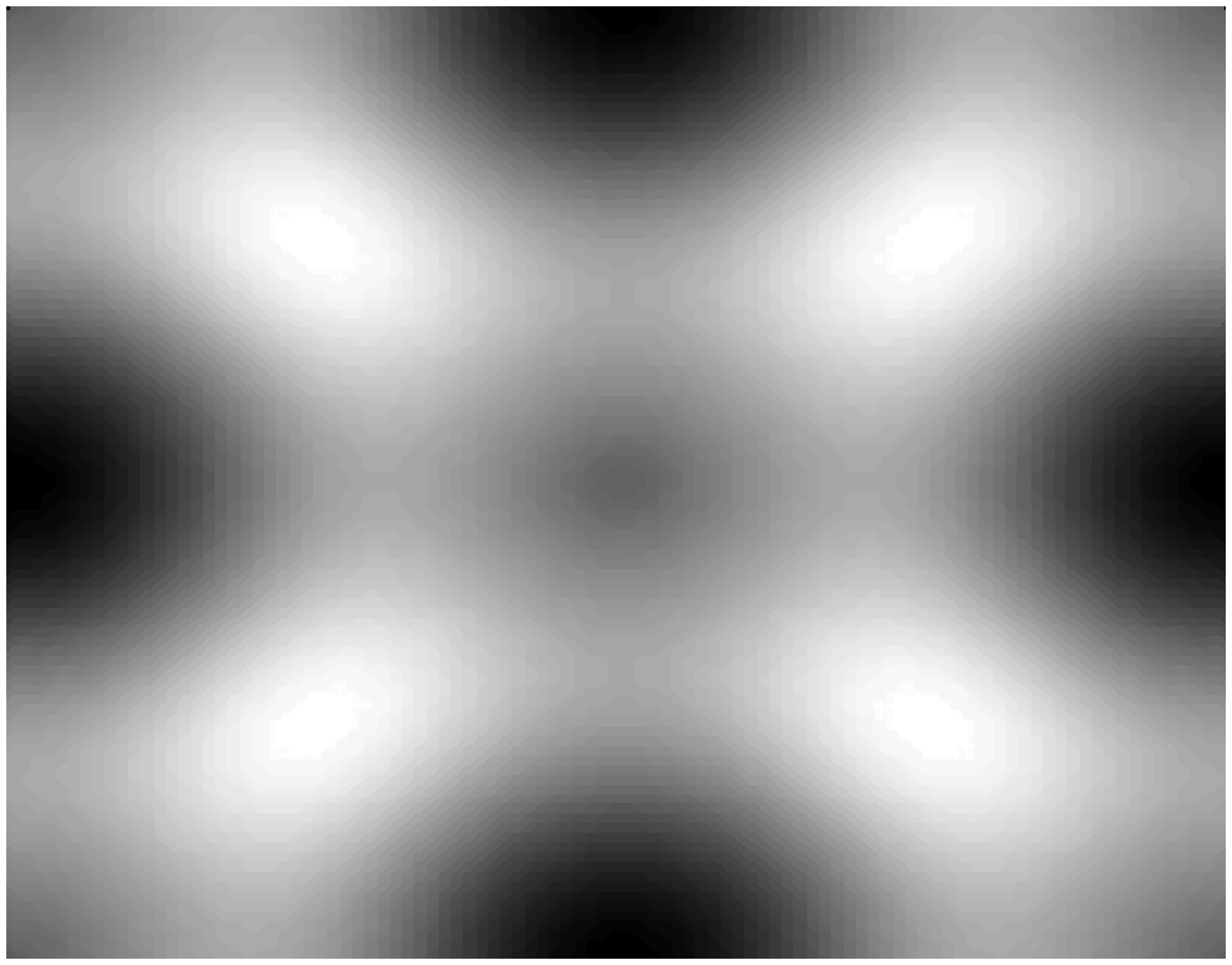}}\\{Fig.\ 13. Charge density profile in the square unit cell in the low stiffness limit
at $\nu=0.98$. }
\end{figure}

\begin{figure}
{\epsfysize2.7in\epsfbox{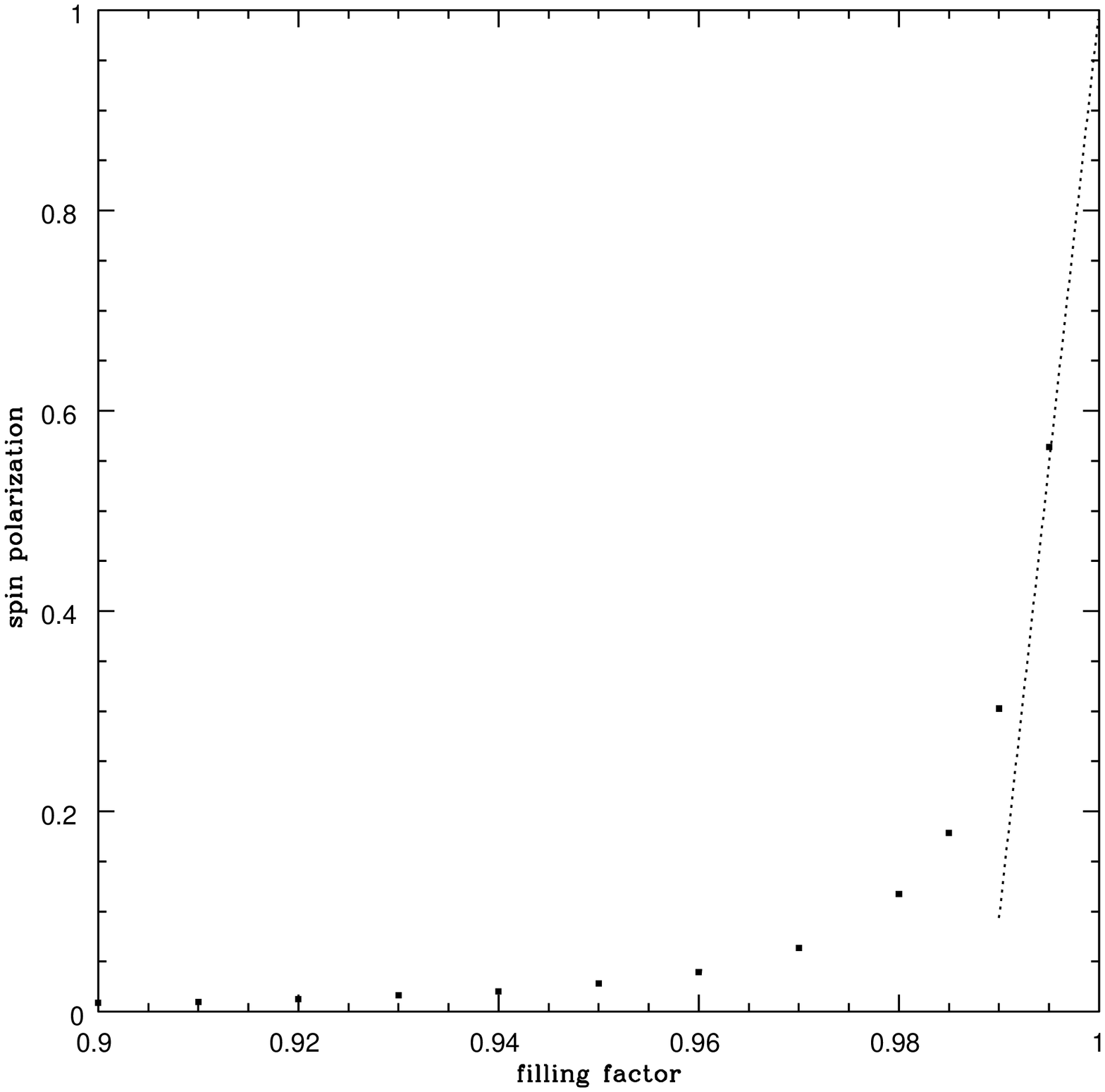}}\\{Fig.\ 14.
Plot of the value of the spin polarization as
a function of the filling factor in the high stiffness regime.
The dotted line is the curve calculated from a dilute lattice of
isolated skyrmions.}
\end{figure}

\begin{figure}
{\epsfysize2.7in\epsfbox{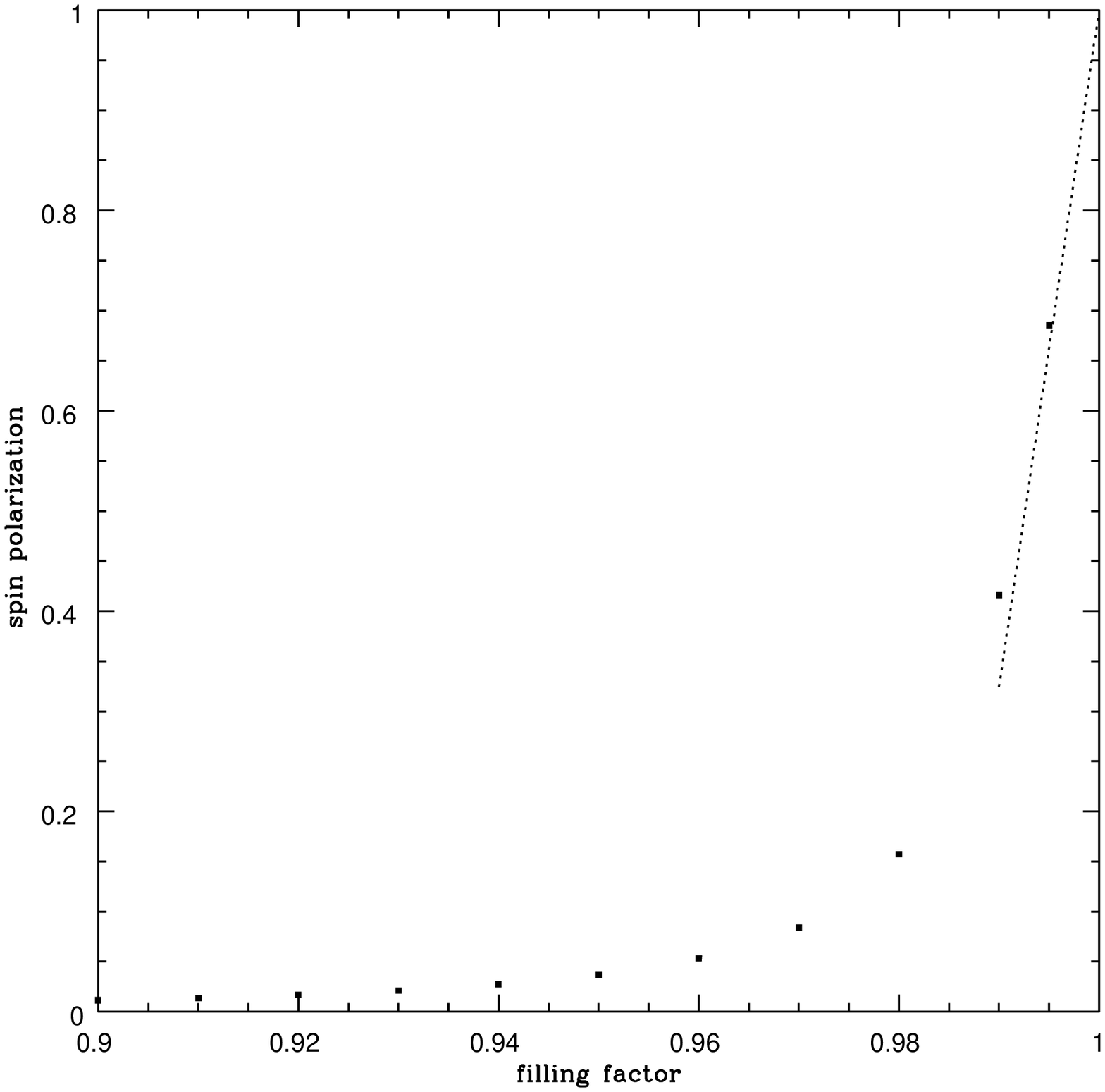}}\\{Fig.\ 15.
Plot of the value of the spin polarization as
a function of the filling factor in the low stiffness regime.
The dotted line is the curve calculated from a dilute lattice of
isolated skyrmions.}
\end{figure}


\begin{thebibliography}{16}
\bibitem{color}
Charge density plots in color are available from the authors on request.
\end{thebibliography}
\end{document}